\documentclass[]{raa}            % referee version: for submission
\usepackage{graphicx,times}             %for PS/EPS graphics inclusion, new
\usepackage{natbib}

\usepackage{amssymb,amsmath,dcolumn}
\usepackage{amsthm,amsmath,amssymb}
\usepackage{mathrsfs}
\usepackage{multirow}

\bibpunct{(}{)}{;}{a}{}{,}
\usepackage{lineno}
\let\oldflalign\flalign \let\oldendflalign\endflalign  \renewenvironment{flalign}    {\linenomathNonumbers\oldflalign}    {\oldendflalign\endlinenomath}
\usepackage[pagebackref=true]{hyperref}

\begin{document}
%\linenumbers

\title{Gravitational Lensing of Gravitational Waves from Astrophysical Sources: Theory, Detection, and Applications
}
%\subtitle{I. Place Your Subtitle Here}

\volnopage{Vol.0 (20xx) No.0, 000--000}      %%preserved for Editor. DOn't remove!
\setcounter{page}{1}          %%starting page, preserved for Editor. DOn't remove!

\author{Zhiwei Chen 
\inst{1,2}
\and Youjun Lu\thanks{Corresponding author: luyj@nao.cas.cn}
\inst{2,1}
}

\institute{National Astronomical Observatories, Chinese Academy of Sciences, 20A Datun Road, Beijing 100101, China; 
\and
School of Astronomy and Space Sciences, University of Chinese Academy of Sciences, 19A Yuquan Road, Beijing 100049, China
}
%\and
%\vs\no
%{\small Received 20xx month day; accepted 20xx month day}

\abstract{ 
Gravitational Waves (GWs) emitted from distant astrophysical sources can be gravitationally lensed by objects or systems encountered along their propagation paths. Strong astrophysical GW sources include inspiralling and merging stellar-mass compact binaries (stellar-mass binary black holes and binary neutron stars), intermediate-mass and supermassive-binary-black-holes. Lenses range from stars, primordial black holes, dark matter halos to galaxies and galaxy clusters. Depending on the ratio of the GW wavelength to the lensing scale, GW lensing can occur in two regimes: geometric-optics, which produces multiple images of a single lensed event with relative time delays and magnifications, and wave-optics, which produces frequency-dependent amplifications and phase shifts in the observed waveform. Lensed GW signals can be identified either by overlap of inferred parameters between event pairs followed by joint Bayesian model comparison or by characteristic frequency-dependent amplification and phase modulation in the waveform that distinguish them from unlensed signals. The detection rates for different classes of lensed GW events are set by the redshift distributions of source populations and of intervening lenses, together with the lenses' mass and spatial distributions, of which the predictions are quite promising for future detection. Once confirmed, lensed GW events will become powerful probes of astrophysical processes, fundamental physics, and cosmology: they can constrain the nature and abundance of dark matter (including compact-object candidates), the mass function and internal structure of lensing galaxies and (sub)halos, the Hubble constant, and other cosmological parameters. In this paper, we provide a concise overview of the gravitational lensing of GWs, covering the theoretical framework, predicted detection rates for lensed inspirals and mergers, search strategies for lensed GW events, and their astrophysical and cosmological applications. We conclude with prospects and future directions for observing and exploiting the lensing of astrophysical GW events.
}
\authorrunning{Chen \& Lu }            %author_head in even pages
\titlerunning{Gravitational Lensing of Gravitational Waves}  % title_head in odd pages
\maketitle

\section{Introduction}

Detection of the gravitational wave (GW) signals from inspiralling and mergers of compact objects [including black holes (BHs) and neutron stars (NSs)] provide a new way to measure their properties and study their origin and nature. The first direct detection of a stellar binary black hole (sBBH) merger (GW150914) by the Laser Interferometer Gravitational wave Observatory (LIGO) confirmed the existence of sBBHs and substantially revised our understanding of sBBH formation \citep[e.g., ][]{2016PhRvL.116m1102A, 2019PhRvX...9c1040A, 2020arXiv201014527A, 2021arXiv211103606T, 2021arXiv211103634T}. The subsequent observations of the binary neutron star (BNS) merger event, GW170817, including its multiwavelength electromagnetic (EM) counterparts not only confirmed the predictions about short $\gamma$-ray burst (sGRB) with afterglow signals and kilonova produced by BNS mergers, but also marked the opening of the new era of multi-messenger astronomy \citep[e.g., ][]{2017ApJ...848L..12A, 2017ApJ...848L..13A, 2017Sci...358.1556C, 2017Natur.551...67P, 2018ApJ...858L..15D, 2019MNRAS.489L..91C, 2019PhRvX...9a1001A}. With the multimessenger (GW + EM) information, one may estimate the mass of the tidal debris of BNS by the observation of the kilonova signals and therefore further constrain the fundamental physics related to the equation of state (EOS) of neutron stars (NSs) \citep[e.g., ][]{2018PhRvL.120z1103M, 2018PhRvL.121i1102D,Gill:2019bvq}. Moreover, with the redshift measurement from the EM signals and the luminosity distance inferred from GW, one can also constrain the Hubble constant to a precision of $\sim 10\%$ with this single object (GW170817) by the ``standard siren'' method \citep[e.g.,][]{1986Natur.323..310S,2017Natur.551...85A, 2019ApJ...876L...7S}. These results underscore the strong prospects for observing compact-binary coalescences (CBCs) via GWs and EMs.

The astrophysical GW sources are quite abundant in the universe and an extremely large number of such sources are expected to be detected in future. Since the first direct detection of GWs by LIGO, more than $300$ stellar CBCs have been detected by LIGO and Virgo, including about three hundreds sBBH mergers, two BNS mergers, and several BHNS mergers \citep[e.g.,][]{2025arXiv250818080T,2025arXiv250818082T}\footnote{see also at the website https://gracedb.ligo.org/}. With further upgrade, the Advanced LIGO plus (LIGO A+) \citep{2020LRR....23....3A} and  LIGO Voyager \citep{2020CQGra..37p5003A} will detect many more in the near future. The third-generation ground-based GW detectors, i.e., Einstein Telescope (ET) \citep{Hild_2011} and Cosmic Explorer (CE) \citep[][]{2019BAAS...51g..35R} are even anticipated to detect more than several tens of thousands of CBCs per year. In addition, the pulsar timing arrays (PTA) \citep[e.g.,][]{2023ApJ...951L...8A,2023A&A...678A..50E,2023ApJ...951L...6R, 2023RAA....23g5024X, 2025MNRAS.536.1467M} are expected to detect the nanohertz GW signals from inspiralling supermassive binary black holes (SMBBHs) with masses $\gtrsim 10^8M_\odot$ \citep[e.g.,][]{2013MNRAS.433L...1S,2024ApJ...974..261C} both collectively as a stochastic background and individually as continuous gravitational wave sources (CGWs). The space-based GW detectors, such as Laser Interferometer Space Antenna (LISA) \citep{2017arXiv170200786A}, Taiji \citep{2020ResPh..1602918L} and Tianqin \citep{2021PTEP.2021eA107M}, are designed to detect the millihertz GW signals from mergers of massive binary black  holes (MBBHs) with masses $\sim 10^4-10^7 M_\odot$ and extreme/intermediate mass-ratio inspirals (EMRIs/IMRIs) \citep[e.g.,][]{2007CQGra..24R.113A,2017PhRvD..95j3012B}. The numbers of such sources in the millihertz and nanohertz bands that may be detected in future can also be substantial.

A small fraction of various types of GW events may be gravitationally lensed by intervening objects/systems, such as stars, galaxies, and galaxy clusters, and they may be detectable and identifiable by future GW detectors. Since the GWs considered for detection (by ground-based GW detectors, space-based GW detectors, and the PTAs) have much longer wavelengths ($\lambda_{\rm GW}$) comparing with those of EM waves from radio to gamma-ray bands, the wave nature of GWs may have to be considered in some cases, especially when $\lambda_{\rm GW} \gg \lambda_{\rm E}$ with $\lambda_{\rm E}$ representing the Einstein radius of the lens (a characteristic scale). In those cases with $\lambda_{\rm GW} \ll \lambda_{\rm E}$, the gravitational lensing of GWs is in the geometric-optics regime, and the lensing effect is similar to that for EM waves (such as lensed QSOs, lensed galaxies, lensed planets, etc.). However, in some cases the GW wavelength can be larger than the Einstein radius of the lens, i.e., $\lambda_{\rm GW} \gtrsim \lambda_{\rm E}$, thus the wave-optics should apply for the lensing and the diffraction is important, which hardly occur for the cases of EM wave lensing. In addition, not like the cases for EM waves that may be easily contaminated by light from foreground galaxies or distorted by interstellar medium (ISM), the GW signals are hardly contaminated and thus encode clean information about physical properties of the sources and the propagation path. Therefore, the lensing of GW signals offers unique a probe for some key astrophysical and cosmological questions \citep[e.g.,][]{2023Univ....9..200G, 2025RSPTA.38340134S,2025RSPTA.38340127M, 2025RSPTA.38340122L}, including the wave nature of GWs, the nature of the dark matter \citep[e.g., ][]{2022PhRvD.106b3018G,2023JCAP...07..007F}, the origin of sBBH mergers \citep[e.g.,][]{2022ApJ...940...17C}, and the cosmological parameters at relatively high redshifts \citep[][]{Liao=2017, 2020MNRAS.498.3395H, zhiwei_na,2023arXiv230106117S}. 

The study of gravitational lensing of GWs thrives rapidly and there has been more than $\sim 800$ related papers in the literature since the first proposal for lensed GW signals by \citet{1996PhRvL..77.2875W}, discussing from both theoretical and observational aspects \citep[see also][for an early review]{2019RPPh...82l6901O,2022ChPhL..39k9801L,Fan:2020sou}. The detection of a large number of GW events by LIGO-Virgo-Kagra (LVK) collaboration imply the detection of the gravitational lensing of GWs will come soon, and ultimately there will be a number of GW events with great details, offering applications to both astrophysics and cosmology. In this paper, we will review the gravitational lensing of GWs from the theoretical perspectives, focusing on the basic theory of GW lensing (Section~\ref{sec:basic}), source and lens population for statistical analysis (Section~\ref{sec:population}), detection rate estimations (Section~\ref{sec:rates}), cosmological applications of lensed events (Section~\ref{sec:application}). In addition, in Section~\ref{sec:search} we will also give a brief introduction on the effort devoted in searching such lensed GW signals in current GWTC data, though still null-detection. Finally, we give the prospects for the gravitational lensing of GWs in Section~\ref{sec:sum}. 

\section{Basic theory}
\label{sec:basic}

The gravitational lensing of GWs refers to the propagation of GWs in the curved spacetime due to an intervening lens (e.g., a galaxy or a cluster of galaxies). Under the standard assumption that the source is located at a cosmological distance, far away from the lens, the incident GW can be approximated as a plane wave. Similar to the EM waves, a gravitational lens can bend and focus GWs. However, due to their long wavelengths, GW lensing sometimes exhibits significant wave diffraction effects, which alters both the waveform's amplitude and phase. This section first reviews the fundamentals of GWs and then introduces the theoretical framework for their lensing. This framework is divided into two distinct regimes according to the ratio ${\lambda_{\rm GW}}/{\lambda_{\rm E}}$: the geometric-optics regime (${\lambda_{\rm GW}}/{\lambda_{\rm E}}\ll 1 $) and the wave-optics regime (${\lambda_{\rm GW}}/{\lambda_{\rm E}}\gtrsim 1$). 

\subsection{GW Basics}

\subsubsection{GW Waveform}
 
The GW waveform is different in different evolution stages of binary mergers, i.e., inspiral, merger, and ring-down. In the merger and ring-down stages, the waveform can only be accurately calculated by using numerical relativity, for their extremely complex spacetime structure, which is computationally expensive. In the inspiral stage, one can analytically obtain the waveform, and such a method also enables an order of magnitude estimation for the waveform in the other two stages \citep{1960Natur.186..535B, 1962RSPSA.269...21B}. Combining the analytical and numerical methods yield more efficiently and accurate estimates and enables construction of GW template banks for CBCs, which are then used to detect and extract signals  from detector data \citep[e.g.,][]{Owen:1995tm,Magee:2019vmb,Cannon:2020qnf,2020ResPh..1602918L,2023LRR....26....2A,2023FrPhy..1864302R,2025arXiv250818081T,Joshi:2025zdu,Joshi:2025nty,2025arXiv250909741N,2025CQGra..42q3001L}. Below, we briefly introduce the generation of GW waveforms. 

Within the linearized theory of gravity, GWs are described as small perturbations propagating in a flat Minkowski background. For a GW source confined to a compact region and located at a cosmological distance from the observer, the GW strain $h_{ij}$ can be expressed via multipole expansion as   
\begin{equation}
h_{ij}=\frac{2G}{c^4 d_{\rm L}}\frac{d^2Q_{ij}}{dt^2},
\end{equation}
with $G$ and $c$ representing the gravitational constant and speed of light, $d_{\rm L}$ representing the luminosity distance of the source, and $Q_{ij}$ representing the quadruple moment of the system, i.e.,  
\begin{equation}
Q_{ij}= \int d^3x\rho(\textbf{x}) \left(x_i x_j-\frac{1}{3}x^2\delta_{ij}\right),
\end{equation}
where $\rho(\textbf{x})$ is the matter density of the source system at position $\textbf{x}$, $\delta_{ij}=1$ when $i=j$ and $0$ when $i\ne j$. The GW radiation power of the system can be expressed as
\begin{equation}
\label{eq:siji}
\frac{dE_{GW}}{dt}=\frac{1}{5}\frac{G}{c^5}\sum_{i,j=1}^{3}\frac{d^3Q_{ij}}{dt^3}\frac{d^3Q_{ij}}{dt^3}.
\end{equation}
Note that assuming the transverse-traceless gauge, i.e., TT gauge, the GW strain tensor can be rewritten as
\begin{equation}
h_{ij}^{\rm TT}(t,z)=\begin{pmatrix}
h_{+} & h_{\times} & 0 \\
h_{\times} & -h_{+} & 0 \\
0 & 0 & 0
\end{pmatrix}_{ij} \cos{[\omega(t-z/c)]},
\end{equation}
where $z$ is the propagation direction of the GW signal, $\omega=2\pi f_{\rm gw}$ with $f_{\rm gw}$ representing the GW frequency in the source frame, and $h_{+}$ and $h_{\times}$ are the plus and cross polarization of the GW, the only two polarizations in Einstein's general relativity (GR). 

As for the circular orbit in the inspiral stage, the GW strain can be expressed directly in terms of the time to coalescence $\tau_{\rm obs}$ in measured by the observer as
\begin{eqnarray}
h_{+}(\tau_{\rm obs})& = & \frac{4}{d_{\rm L}}\left(\frac{GM_{{\rm c},z}}{c^2}\right)^{\frac{5}{3}}\left(\frac{\pi f_{\rm gw}^{\rm (obs)}(\tau_{\rm obs})}{c}\right)^{\frac{2}{3}}\left(\frac{1+\cos^2{\iota}}{2}\right)\cos{[\Phi(\tau_{\rm obs})]}, \\
h_{\times}(\tau_{\rm obs}) & = & \frac{4}{d_{\rm L}}\left(\frac{GM_{{\rm c},z}}{c^2}\right)^{\frac{5}{3}}\left(\frac{\pi f_{\rm gw}^{\rm (obs)}(\tau_{\rm obs})}{c}\right)^{\frac{2}{3}}\cos{\iota}\sin{[\Phi(\tau_{\rm obs})]}.
\end{eqnarray}
Here $\iota$ denotes the inclination angle between the binary-orbital normal and the line of sight; $M_{{\rm c},z}(z)=(1+z)M_{\rm c}$ is the redshifted chirp mass, where $M_{\rm c}=M_1^{3/5}M_2^{3/5}/(M_1+M_2)^{1/5}$, and $M_1$ and $M_2$ are the primary and secondary masses; $\Phi(\tau_{\rm obs})$ is the GW phase in the observer's frame described as
\begin{equation}
\Phi(\tau_{\rm obs})=-2\left(\frac{5GM_{{\rm c},z}}{c^3}\right)^{-\frac{5}{8}}\tau^{\frac{5}{8}}_{\rm obs}+\Phi_0;
\end{equation}
and $f^{\rm obs}_{\rm gw}=f_{\rm gw}/(1+z)$ is the GW frequency in the observer's frame which can be expressed as
\begin{equation}
f_{\rm gw}^{\rm obs}(\tau_{\rm obs})=\frac{1}{\pi}\left(\frac{5}{256}\frac{1}{\tau_{\rm obs}}\right)^{\frac{3}{8}}\left(\frac{GM_{{\rm c},z}}{c^3}\right)^{-\frac{5}{8}}.
\end{equation}
For a binary on a circular orbit, the GW frequency is twice the orbital frequency, ($f_{\rm gw}=2f_0$), where $f_0=(Gm/a^3)^{1/2}$ and $a$ is the semi-major axis. 

With the above equations, one can estimate the inspiral GW waveform $h(\tau_{\rm obs})$ received by the  observer as
\begin{equation}
h=F_{+}h_{+}+F_{\times}h_{\times}, 
\end{equation}
where $F_{+}$ and $F_{\times}$ are the antenna pattern function of the GW detector. In practical GW data analysis, we usually Fourier-transform the signal to the frequency domain for convenience, which is equivalent to working in the time-domain. An example for the GW waveform can be given by the first direct detection of GWs, GW150914, with component masses of $\sim 36M_\odot$ and $\sim 29 M_{\odot}$ at a luminosity distance of $\sim 410$\,Mpc, its strain amplitude at Earth is of the order $10^{-21}$, showcasing the remarkable measurement capabilities required for the GW detection \citep{2016PhRvL.116x1102A}. 

The above expressions are only approximations under the Newtonian limit, which cannot be applied to late time of inspiral, merger, and ring-down stages under extremely strong gravitational field. Generally speaking, the waveform in the later inspiral stage is often calculated via the post-Newtonian expansion \citep[e.g.,][]{1980RvMP...52..299T, 2000PhRvD..62h4011D,2023arXiv230603797P, 2014LRR....17....2B}, in which the higher order nonlinear terms of the field equation are considered. Contributions scaling as $(v/c)^{2n}$ are labeled as the $n\rm PN$ term in the post-Newtonian expansion. Extending the series to high-order PN term is time-consuming and tedious, and to date waveform models consider such terms only up to $4.5\rm PN$ in the literature \citep{2017PhRvD..95l4001M}. As for the merger process, the GW waveform may be calculated by using numerical relativity \citep[e.g.,][]{2010nure.book.....B} through $3+1$ decomposition adopting the Arnowitt-Deser-Misner (ADM) formalism \citep{2008GReGr..40.1997A}, in which the spacetime is split with several space-like hyperspace. When it comes to the ring-down process, the GW waveform can be modeled with the BH perturbation theory \citep[e.g.,][]{2008PhRvD..77j4017A}, by solving the Teukolsky and Zerilli equations \citep{1972PhRvL..29.1114T, 1972PhRvD...5.2419P}. Here we do not intend to overview the details of the GW calculations for these stages, which can be seen in the classical textbook of \citet{GWbook1} and \citet{GWbook2} for interested readers. However, we noticed that it is not possible to calculate the waveform across all the allowed parameter space, as GW waveform calculations by the above methods are indeed time-consuming. Therefore, in practical searches and data analysis, the GW template is normally estimated by data-driven phenomenological waveform model, which balance computational efficiency with physical fidelity. These models parameterize the signal with a small number of physical parameters (e.g., masses, spins), enabling simultaneous parameter estimation. There are three main phenomenological model series, i.e., IMRPhenom Series \citep[e.g.,][]{2023ascl.soft07019P, 2019PhRvD.100b4059K, 2017PhRvD..96l1501D}, SEOBNR Series \citep[e.g.,][]{2023PhRvD.108l4037R, 2015PhRvD..92j2001K}, and NRSur Models \citep[e.g.,][]{Islam:2023zzj, 2024arXiv241206946G}, which have been widely used in GW data analysis. Currently, these models have been integrated within the main GW data analysis tool package \texttt{PyCBC}
\footnote{https://pycbc.org/}
\citep{2019PASP..131b4503B} and readers can generate GW waveforms for various sources with various parameters conveniently via this package.  

\subsubsection{GW Data Analysis}
\label{sec:GWdataana}

The strain of the GW signal is extremely small and may be substantially lower than the noise of the GW detector. Therefore, it is important to process the output of the detector adequately, in case no real signals are missed. In principle, the output of a GW detector can be viewed as the linear superposition of the noise $n(t)$ and the signal $h(t)$, i.e.,
\begin{equation}
s(t)=h(t)+n(t).
\end{equation}
Note that $n(t)$ accounts for all possible noises in the GW detection, including the seismic, quantum, thermal noises, and so on. We can always assume these noises to be static and Gaussian, then their Fourier components can be described by the single-sided power spectral density (PSD) as
\begin{equation}
%
%S_{\rm n}(f) \equiv 
\left<n(f)n(f^{\prime})\right>=\frac{1}{2}\delta(f-f^{\prime})S_{\rm n}(f),
\end{equation}
where the bracket symbol $\left<\cdots \right>$ marks the ensemble average. Here we comment that the above definition of $S_{\rm n}(f)$ is valid only for $n(t)$ that can have Fourier component(s), which is not always satisfied. If $n(t)$ does not have Fourier components, we can only define the PSD by the Fourier transform of the auto-correlation function of the noise with Wiener-Khintchin relation \citep{GWbook1,GWbook2}.

There are two key aspects in the GW data analysis: (1) when can we claim a GW detection? (2) how can we extract parameters of the GW event from the waveform? After defining the PSD of the GW detector, we can answer these two questions by using the matched filtering method, the main method of searching and identifying GW signals in observational data. The principle of the matched filtering is to maximize the signal-to-noise ratio (SNR or S/N) by finding the optimal way to match the template to the data, and the S/N is defined as
\begin{equation}
\varrho
=\frac{\int_{-\infty}^{\infty}df{h}(f){K}^{*}(f)}{\left[\int_{-\infty}^{\infty}df(1/2)S_{\rm n}(f)|{K}(f)|^{2}\right]^{1/2}}.
\end{equation}
The function $K(f)$ is the filter function and it is meant to maximize S/N for a given $h(t)$. According to the Schwartz inequality, the optimal filter is
\begin{equation}
K(f)={\rm const} \times \frac{h(f)}{S_{\rm n}(f)},
\end{equation}
where the constant is arbitrary. Then, the optimal value of S/N is given by
\begin{equation}
%
%\left(\frac{S}{N}\right)^2
\varrho^2
=4\int_0^{\infty}df\frac{|h(f)|^2}{S_{\rm n}(f)}.
\end{equation}
The above equation is completely general and independent of the form of $h(f)$.  The above expression for S/N is valid for single detector. For a detector network composed of $n$ detectors ($n=1$ for a single detector), it is useful to define the whitened GW data sets of the network 
\begin{equation}
\hat{\mathbf{d}}(f)=\left(\frac{A_1(f)h_1(f)}{\sqrt{S_{{\rm n},1}(f)}},\frac{A_2(f)h_2(f)}{\sqrt{S_{{\rm n},2}(f)}}, \cdots, \frac{A_{m}(f)h_{m}(f)}{\sqrt{S_{{\rm n},m}(f)}}\right),
\end{equation}
where $A_{m}=e^{-2\pi i f((\hat{r}_{m}-\hat{r}_{1})\cdot \hat{n}_{\rm GW})}$ is the phase transfer function, $\hat{r}_{m}$ is the location of the $m$-th detector, $\hat{n}_{\rm GW}$ is the unit direction vector of the GW source, and $S_{{\rm n},m}$ denotes the one-sided power spectrum of the corresponding $m$-th GW detector \citep[e.g.,][]{2010PhRvD..81h2001W}. 
Then the optimal squared S/N is given by
\begin{equation}
\varrho_{\rm GW}^2=\left\langle\hat{\mathbf{d}}(f)|\hat{\mathbf{d}}(f)\right\rangle,
\label{eq:SNR}
\end{equation}
where the angular bracket denotes an inner product. For any two vector functions $\hat{\mathbf{a}}(f)$ and $\hat{\mathbf{b}}(f)$, this inner product is defined as
\begin{equation}
\langle \hat{\mathbf{a}}(f)| \hat{\mathbf{b}}(f)\rangle=2\sum_{j} \int_{f_{\text {min }}}^{f_{\text {max }}}\left\{a_j(f) b_j^{*}(f)+a_j^{*}(f) b_j(f)\right\} d f ,
\end{equation}
where $j$ denotes the $j$-th component of the vector, ${f_{\rm min}}$ and ${f_{\rm max}}$ are the lower and upper frequency limits of the GW waveforms. Note that in practice a detection threshold is imposed on the matched-filer S/N ratio, typically denoted as $\varrho_0$ and commonly set to $8$; this choice of $\varrho_0$ keeps the false alarm low so that the expected number of false alarms per observing run is on the order of a few or less. More details on this point can be seen in Section~7.4 of \citet{GWbook1}.

As for the parameter estimation in the GW data analysis, the full problem can be stated as: how do we reconstruct the most probable values for the source parameters, and how we compute the uncertainties of these parameter estimates? This problem is naturally Bayesian and the posterior of the probability is 
\begin{equation}
\label{eq:post}
p(\chi_t|s)=\mathcal{N}p^{(0)}(\chi_t)\exp\left\{\left<h_t|s\right>-\frac{1}{2}\left<h_t|h_t\right>\right\},
\end{equation}
where $\chi_{t}$ is the unknown true value of parameter $\chi$, $h_{t}$ is the strain with parameter $\chi_t$, i.e., $h(\chi_{t})$ and $p^{(0)}(\chi_t)$ is the prior probability. Note that the above posterior is only valid for Gaussian and Stationary noises, so that the Gaussian likelihood emerged. Full parameter estimation for GW signals can be conducted with Equation~\eqref{eq:post} by Monte Carlo method \citep[e.g.,][]{2019ApJS..241...27A,2020MNRAS.499.3295R}, which is computationally demanding. An approximate way to estimate the uncertainties of these parameters is to calculate the Fisher information matrix, which is actually the linear term of the expansion of the above posterior,
\begin{equation}
p(\chi|s)\propto \exp\left\{  -\frac{1}{2}\Gamma_{jk}\Delta\chi_{j}\Delta\chi_{k}  \right\},
\end{equation}
at most probable values $\hat{\chi}_{\rm B}^{i}(s)$, 
\begin{equation}
\hat{\chi}_{\rm B}^{i}(s)=\int d\chi\chi^{i}p(\chi|s),
\end{equation}
where $\Delta\chi_{i}=\chi_{i}-\hat{\chi}_{\rm B}^{i}$. In the specific case of large $S/N$ and GW detector network, the fisher matrix can be estimated by $\Gamma_{jk}$, 
\begin{equation}
\Gamma_{j k}=\left\langle\partial_{j} {\hat{ \mathbf{d}}(f)} | \partial_{k} \hat{\mathbf{d}}(f)\right\rangle,
\end{equation}
where $\partial_j$ and $\partial_k$ denote the partial derivative with respect to the $j$-th and $k$-th parameter, respectively. Then the expectation value of the errors $\delta\chi_{i}$ are given by:
\begin{equation}
{\rm Cov}(\Delta\chi_{j},\Delta\chi_{k})=(\Gamma_{jk})^{-1}.
\end{equation}
As an example, we can estimate the localization precision of GW sources in the sky
\begin{equation}
\Delta \Omega=2 \pi\left|\sin \vartheta_{\rm s}\right| \sqrt{\left\langle\Delta {\vartheta}_{\rm s}^{2}\right\rangle\left\langle\Delta \phi_{\rm s}^{2}\right\rangle-\left\langle\Delta \vartheta_{\rm s} \Delta \phi_{\rm s}\right\rangle^{2}},
\label{eq:omega}
\end{equation}
where $\Delta \vartheta_{\rm s}$ and $\Delta \phi_{\rm s}$ are the standard deviations obtained from the covariance matrix. As a straight forward magnitude estimation, the median localization precision of sBBH mergers detected by the third-generation GW detectors (such as ET and CE) will be about $\sim 1 {\rm deg^2}$, which is still large for EM telescope follow-up. More detailed descriptions on the estimation of the GW localization precision for CBCs can be seen in \citet{2018PhRvD..97f4031Z} and \citet{2022arXiv220312586P}. In the Fisher information matrix approach we neglect nonlinear signal terms \citep{PhysRevD.77.042001}, an approximation that holds only when the S/N is sufficiently high. It has been shown that for systems with sufficiently high S/Ns ($\ge 10$), parameter estimates from the Fisher information matrix method agree well with those from full Bayesian analysis \citep[e.g.,][]{2013PhRvD..88h4013R, PhysRevD.89.042004}. However, one should be cautious when applying the Fisher information matrix method to GW events with S/N substantially less than $10$.

\subsubsection{Current and Future GW detectors}

Here we list some current and future GW detectors designed for detecting GW sources in different frequency bands for references. 

\begin{itemize}
\item[(1)] \textbf{High-frequency ($\rm 10-100$\,Hz)}: The main sources in this band are CBCs (i.e., sBBH, BHNS, and BNS mergers), core-collapse supernovae, and rotating asymmetric NSs. The corresponding detectors are ground-based laser interferometer detectors, including LIGO \citep[e.g., ][]{2016PhRvL.116m1102A}, Virgo \citep{1990NIMPA.289..518B}, KAGRA \citep{2019NatAs...3...35K}, Indigo \citep[][]{2023arXiv230107522U}, and the third-generation GW detectors, such as CE \citep{2019BAAS...51g..35R} and ET \citep{Hild_2011}.
\item[(2)] \textbf{Middle-frequency ($\rm 0.1-10$\,Hz)}: The main sources are intermediate mass BBH mergers and inspiraling binary compact objects. The corresponding detectors are space-based decihertz laser interferometer detectors, such as the proposed DECIGO \citep{2011CQGra..28i4011K}, BBO \citep{bbo} and AMIGO \citep{2018EPJWC.16801004N}.
\item[(3)] \textbf{Low-frequency ($10^{-4}-0.1$\,Hz)}: The main sources are mergers of massive binary black holes (MBBHs) with mass $\sim 10^{4}-10^7M_{\odot}$, and extreme/intermediate-mass-ratio-inspirals (EMRIs/IMRIs), inspiraling double white dwarfs (DWDs). The corresponding detectors are space-based millihertz laser interferometer detectors, such as future LISA \citep{2017arXiv170200786A}, Taiji \citep{2020ResPh..1602918L}, and Tianqin \citep{2021PTEP.2021eA107M}.
\item[(4)] \textbf{Very low-frequency ($\rm 10^{-9}-10^{-7}$\,Hz)}: The loud astrophysical GW sources in this band are inspiraling SMBBHs with masses $\gtrsim 10^{8} M_{\odot}$. Some physical processes occurred in the early universe can also produce stochastic GW background in this band \citep[see][for a thorough comparison]{2024PhRvD.109b3522E}, such as cosmic (super) strings \citep{1995RPPh...58..477H}, cosmological first-order phase transition \citep{PhysRevD.30.272} and primodial scalar-induced GW \citep{10.1143/PTP.54.730}, etc. The corresponding detectors are pulsar time arrays (PTAs), including Chinese PTA (CPTA), Parkes PTA (PPTA), North American Nanohertz Observatory for Gravitational waves (NANOGrav), European PTA (EPTA, with Indian PTA, denoted as InPTA), Meerkat PTA (MPTA), and future Square Kilometer Array PTA (SKA-PTA) and next generation Very Large Array (ngVLA) \citep[e.g.,][]{2022PASA...39...53T,2023ApJ...951L...8A, 2023A&A...678A..50E, 2023ApJ...951L...6R, 2023RAA....23g5024X, 2025MNRAS.536.1467M}. 
\item[(5)] \textbf{Extreme low-frequency ($\rm 10^{-18}Hz-10^{-15}Hz$)}: In this band, the main target is the primordial stochastic GW background resulting from inflation and it may be detected by CMB polarization measurements.
\end{itemize}

\subsection{Lensing Basics}

\subsubsection{Geometric Optics}

\begin{figure}
\centering
\includegraphics[width=0.82\columnwidth]{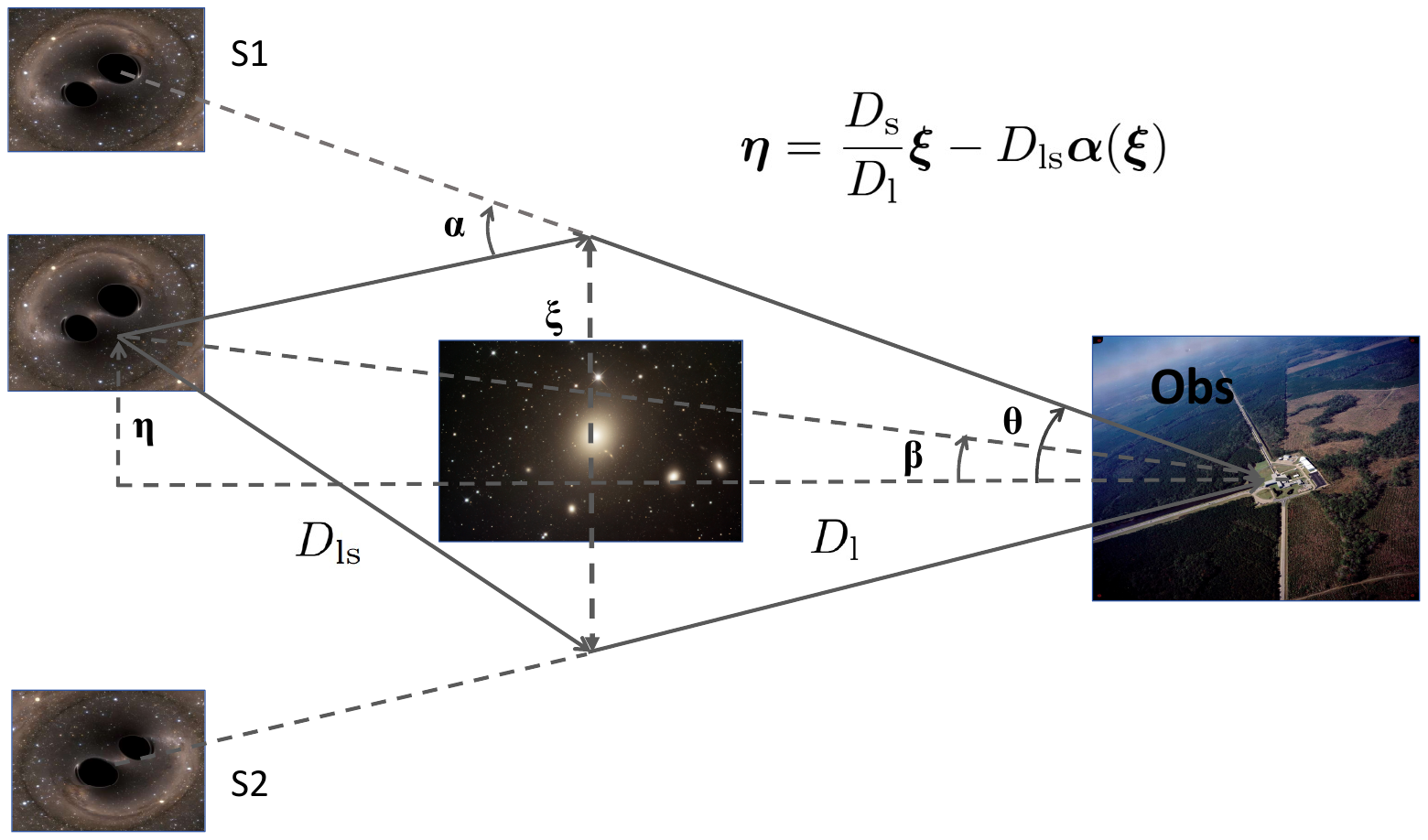}
\caption{A schematic diagram to illustrate the gravitational lensing of gravitational waves in the geometric-optics regime. This diagram shows the gravitational waves from a merging sBBH event is lensed by an intervening galaxy and split into two images $S1$ and $S2$. The geometry is simply described by the lens equation under thin lens approximation, as shown in the figure.}
\label{fig:f1}
\end{figure}

\begin{figure}
\centering
\includegraphics[width=0.82\columnwidth]{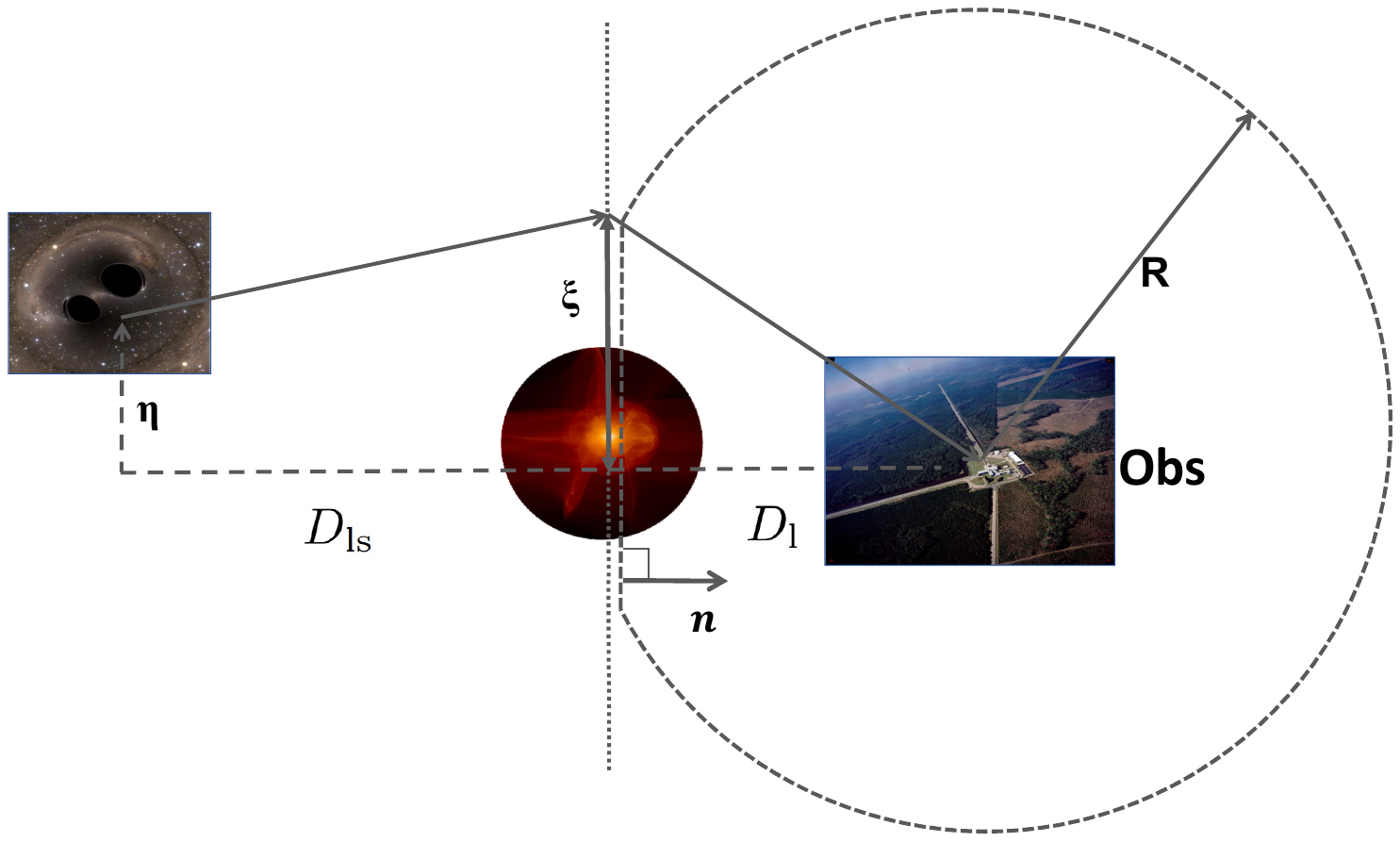}
\caption{
A schematic diagram to illustrate the gravitational lensing of gravitational waves in the wave-optics regime. This diagram shows the gravitational waves emitted from a merging sBBH event is diffractively lensed by a dark matter minihalo.
}
\label{fig:f2}
\end{figure}

In the geometric-optics regime, i.e., ${\lambda_{\rm GW}}/{\lambda_{\rm E}}\ll 1$, the diffraction effect is not important. Assuming the graviton to be massless, GW propagates along the null geodesics in the space-time. Here we only sketch out the classical equations under the geometrically-thin lens approximation. More detailed descriptions can be found in \citet{Schneider=2006}. 

Suppose that a GW event is lensed by an intervening galaxy as lens in the geometric-optics regime, for illustration, producing two images of the GW signals. Figure~\ref{fig:f1} shows a schematic diagram for this lens system. The deflection angle of GW by an arbitrary mass density distribution is given by
\begin{equation}
\boldsymbol{\alpha(\xi)}=\frac{4G}{c^2}\int d^2\xi^{\prime}\Sigma(\boldsymbol{\xi^{\prime}})\frac{\boldsymbol{\xi-\xi^{\prime}}}{|\boldsymbol{\xi-\xi^{\prime}}|^2},
\end{equation}
where $\boldsymbol{\xi}$ denotes the coordinate in the lens plane and $\Sigma(\boldsymbol{\xi})$ is the surface mass density. As sketched in Figure~\ref{fig:f1}, the source and lens planes are defined as planes perpendicular to the optical axis from the observer to the lens. We denote the angular diameter distances from the source to the observer, from the lens to the observer, and from the source to the lens as $D_{\rm s}$, $D_{\rm l}$, and $D_{\rm ls}$, respectively. According to the simple geometry shown Figure~\ref{fig:f1}, we have
\begin{equation}
\boldsymbol{\eta}=\frac{D_{\rm s}}{D_{\rm l}}\boldsymbol{\xi}-D_{\rm ls}\boldsymbol{\alpha}(\boldsymbol{\xi}),
\end{equation}
where $\boldsymbol{\eta}$ is the coordinate of the source in the source plane. By introducing angular coordinates, i.e.,  $\boldsymbol{\beta} = \boldsymbol{\eta} /D_{\rm s}$ and $\boldsymbol{\theta} = \boldsymbol{\xi}/ D_{\rm l}$, one can define the scaled lens equation as
\begin{equation}
\boldsymbol{\beta}=\boldsymbol{\theta}-\frac{D_{\rm ls}}{D_{\rm s}}\boldsymbol{\alpha}(D_{\rm l}\boldsymbol{\theta})=\boldsymbol{\theta}-{\boldsymbol{\Tilde\alpha}(\boldsymbol{\theta})},
\label{eq:lens}
\end{equation}
where ${\boldsymbol{\Tilde\alpha}(\boldsymbol{\theta})}$ is the scaled deflection angle. The scaled lens equation above shows that a source with a true position $\boldsymbol{\beta}$ can be seen by the observer at an angular position of $\boldsymbol{\theta}$. The solution of $\boldsymbol{\theta}$ from the above lens equation may be not necessarily unique. In another words, the lensing effect can lead to multiple images of a single GW signal.

For convenience of calculation, we may further rewrite the scaled deflection angle $\boldsymbol{\Tilde\alpha}(\boldsymbol{\theta})$ in terms of the surface mass density as 
\begin{equation}
\boldsymbol{\Tilde\alpha}(\boldsymbol{\theta})=\frac{1}{\pi}\int d^2\boldsymbol{\theta^{\prime}} \kappa(\boldsymbol{\theta^{\prime}})\frac{\boldsymbol{\theta-\theta^{\prime}}}{|\boldsymbol{\theta-\theta^{\prime}}|^2},
\end{equation}
where $\kappa(\boldsymbol{\theta})$ is the dimensionless surface mass density or convergence described as
\begin{equation}
\kappa(\boldsymbol{\theta})=\left(\frac{c^2}{4\pi G}\frac{D_{\rm s}}{D_{\rm l}D_{\rm ls}}\right)^{-1}\Sigma(D_{\rm l}\boldsymbol{\theta}).
\end{equation}
The scaled deflection angle can be expressed as the gradient of a deflection potential, i.e.,
\begin{equation}
\label{eq:potential}
\psi(\boldsymbol{\theta})= \frac{1}{\pi}\int d^2\boldsymbol{\theta^{\prime}} \kappa(\boldsymbol{\theta^{\prime}})\ln|\boldsymbol{\theta-\theta^{\prime}}|,
\end{equation}
where $\boldsymbol{\Tilde\alpha}(\boldsymbol{\theta}) =\nabla \psi$, and $\psi$ can be related with $\kappa$ by the Poisson equation
\begin{equation}
\nabla^2\psi=2\kappa,
\end{equation}
which is similar to that in the standard three-dimensional gravity potential theory. Here we refer the readers to several popular softwares useful for solving the above lens equations numerically for various lens models such as \texttt{Lenstronomy} \footnote{https://lenstronomy.readthedocs.io/en/latest/} \citep{2018PDU....22..189B}, \texttt{PyAutoLens} \footnote{https://pyautolens.readthedocs.io/en/latest/}
\citep{2015MNRAS.452.2940N} and \texttt{Lenstool}
\footnote{https://projets.lam.fr/projects/lenstool/wiki} \citep{2011ascl.soft02004K}. 

The energy flux of GW per surface unit at source position $\boldsymbol{\beta}$ can be estimated by
\begin{equation}
I^{(\rm s)}_{f}(\boldsymbol{\beta})=\frac{dE(\boldsymbol{\beta})}{dAdf}=\frac{\pi c^3}{2 G}f^2 (|\Tilde{h}_{+}(f)|^2+|\Tilde{h}_{\times}(f)|^2),
\end{equation}
where $\Tilde{h}_{+}(f)$ and $\Tilde{h}_{\times}(f)$ are the Fourier transform of GW strain in the frequency-domain. Due to conservation of GW energy, the observed energy in the lens plane $I^{(\ell)}_{f}(\boldsymbol{\theta})$ is equal to the $I^{(\rm s)}_{f}(\boldsymbol{\beta})$. Therefore, the flux ratio between the observed and emitted GWs can be estimated as 
\begin{equation}
\mu^{\ell}=\frac{\iint I^{(\ell)}_{f}(\boldsymbol {\theta})d^2\boldsymbol{\theta}}{\iint I^{(\rm s)}_{f}(\boldsymbol {\beta})d^2\boldsymbol{\beta}}=\frac{\iint I^{(\ell)}_{f}(\boldsymbol {\theta})d^2\boldsymbol{\theta}}{\iint I^{(\ell)}_{f}(\boldsymbol {\theta})|\frac{\partial \boldsymbol{\beta}}{\partial \boldsymbol{\theta}}|d^2\boldsymbol{\theta}},
\end{equation}
where $\left|\frac{\partial \boldsymbol{\beta}}{\partial \boldsymbol{\theta}}\right|$ is the Jacobian matrix determined by the scaled lens equation. If the size of the GW source is infinitesimal small, the value of $|\frac{\partial \boldsymbol{\beta}}{\partial \boldsymbol{\theta}}|$ is related with the second derivative of the deflection potential $\psi$. Therefore, one can find that the amplitude of GW strain observed in the lens plane is magnified by a factor of $\sqrt{|\mu|}$, i.e.,  $\Tilde{h}^{(\ell)}_{+,\times}(f)=\sqrt{|\mu|} \Tilde{h}_{+,\times}(f)$. Note that in the geometric-optics regime, $\mu$ is independent of the GW frequency $f$, and thus we have similar expression in the time-domain, i.e., $h^{(\ell)}_{+,\times}(t)=\sqrt{|\mu|} h_{+,\times}(t)$. Similar to the lensing of EM waves, there will be time-delays between the multiple images of the GW signal. For example, as for the first and second images, the time-delay $\Delta \tau_{12}$ can be calculated by  
\begin{equation}
\Delta \tau_{12}= T_1-T_2=
\frac{D_{\rm l} D_{\rm s}}{c D_{\rm ls}}\left[\frac{1}{2}(\boldsymbol{\theta_1}-\boldsymbol{\beta})^2-\frac{1}{2}(\boldsymbol{\theta_2}-\boldsymbol{\beta})^2-(\psi(\boldsymbol{\theta_1})-\psi(\boldsymbol{\theta_2}))\right],
\end{equation}
where the first two terms in the right side of the equation represent the geometric time-delay due to the bending of the path and the second term represents the time-delay due to the gravitational potential, i.e., the Shapiro delay. This time delay will add a steady phase shift to the lensed GW signal, i.e., $h^{(\ell)}_{+,\times}(t)=\sqrt{|\mu^{\ell}|} h_{+,\times}(t-T_{i})$ or equivalently $\Tilde{h}^{(\ell)}_{+,\times}(f)=\sqrt{|\mu^{\ell}|} \Tilde{h}_{+,\times}(f)e^{-j 2\pi f T_{i}}$, where $j=\sqrt{-1}$. In addition, we note that the polarization planes of GWs will also be rotated by the lensing (i.e., GW Faraday rotation), though this effect is rather weak and can be ignored \citep[see][for a theoretical verification]{2019PhRvD.100f4028H}. 

This suggests that the gravitational lensing in geometric-optic regime adjusts the GW waveform mainly in two aspects: 1) the amplitude is magnified by $\sqrt{|\mu^{\ell}|}$; 2) the phase is retarded by $2\pi f T_{i}$.  

\subsubsection{Wave Optics}

In the wave-optics regime, i.e., ${\lambda_{\rm GW}}/{\lambda_{\rm E}}\gtrsim 1 $, the diffraction of the GWs becomes important, which may also lead to changes of the waveform in both amplitude and phase. Here we briefly review the basic theory of the gravitational lensing of GWs in the wave-optics regime, focusing on the typical weak-field case.  Extensions to strong-field wave optics can be found in \citet{chan2025} and \citet{2026PhRvD.113h4056S}.
Figure~\ref{fig:f2} shows a schematic diagram for the lensing of a GW event by a (dark matter) minihalo in the wave-optics regime. More detailed discussions can be seen in \citet{Takahashi_2003}. 

Given the potential of the lens object/system ${U(\boldsymbol{r})}\ll1$, the metric can be expressed as
\begin{equation}
ds^2=-(1+2U)dt^2+(1-2U)d\boldsymbol{r}^2=g^{(\rm B)}_{\mu\nu}dx^{\mu}dx^{\nu},
\end{equation}
where $g^{(\rm B)}_{\mu\nu}$ is the background metric tensor. Notably, this expression is valid for static lens, while the case of spinning-lens can be treated by adding an angular-momentum related term as in \citet{Prabhu:2025elp}. Consider a linear perturbation $h_{\mu\nu}$ in the background metric $g^{(\rm B)}_{\mu\nu}$, the metric is then
\begin{equation}
g_{\mu\nu}=g^{(\rm B)}_{\mu\nu}+h_{\mu\nu}.
\end{equation}
Adopting the TT gauge, we have the equation of motion for the GW strain as
\begin{equation}
h_{\mu\nu;\alpha}^{;\alpha}=0. 
\label{eq:motion}
\end{equation}
Note that the above equation is valid only for $\lambda_{\rm GW}$ much smaller than the typical radius of the curvature of the background. Adopting the eikonal approximation, the GW can be expressed as a form of scalar wave, i.e., 
\begin{equation}
h_{\mu\nu}=\phi e_{\mu\nu},
\label{eq:ek}
\end{equation}
where $\phi$ is a scalar wave, $e_{\mu\nu}$ is the GW polarization tensor and it is transported along the null geodesics. For the gravitational lensing of GWs, $U$ is much smaller than 1, and thus $e_{\mu\nu}$ is almost a constant tensor. Therefore, we only need to consider the change of the scalar wave $\phi$. Inserting Equation~\eqref{eq:ek} into Equation~\eqref{eq:motion}, we obtain the propagation equation of the scalar wave 
\begin{equation}
\partial_{\mu}\left(\sqrt{-g^{(\rm B)}}g^{(\rm B)\mu\nu}\partial_{\nu} \phi\right)=0,
\end{equation}
where $g^{(\rm B)}$ is the determinant of the background metric. For the scalar wave in the frequency domain, we then obtain the so-called Kirchhoff diffraction equation
\begin{equation}
(\nabla^2+\omega^2)\Tilde{\phi}=4\omega^2U \Tilde{\phi},
\end{equation}
where $\Tilde{\phi}(f,\boldsymbol{r})$ is the scalar wave in the frequency domain. Here we comment that there will be no multiple images of GW in the wave-optics regime. The actual propagation path of the diffracted GWs is the Feynman path integral of all possible null-geodesics. 

\begin{figure}
\centering
\includegraphics[width=0.9\columnwidth]{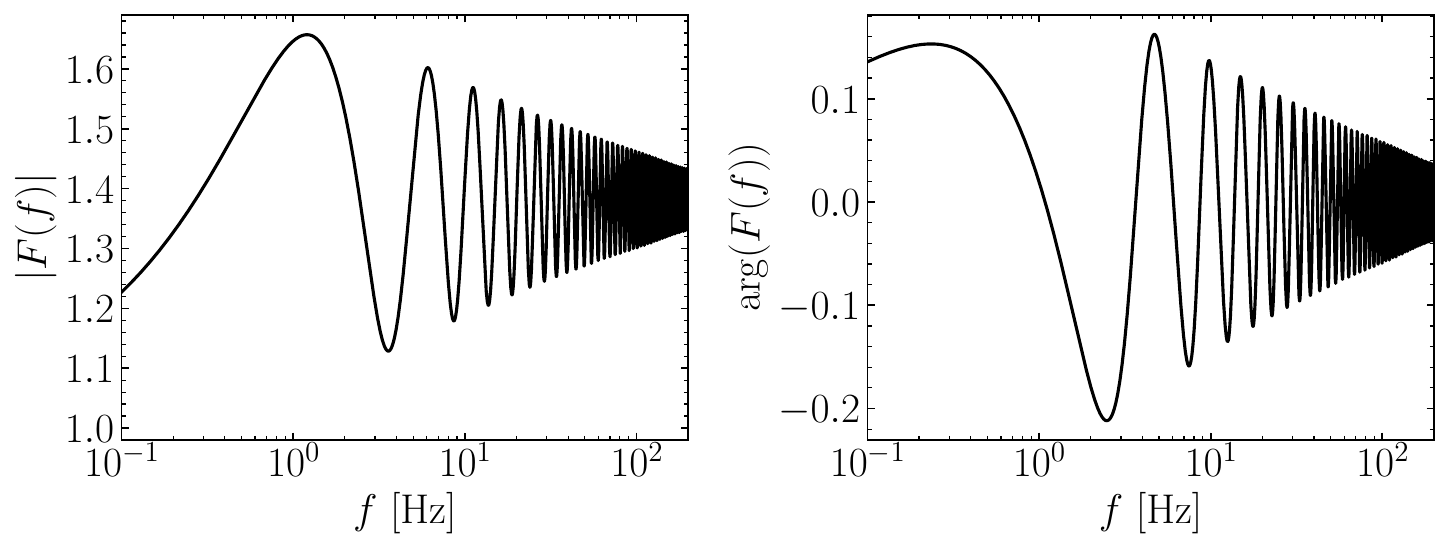}
\caption{
The amplification factor $F(f)$ (left: absolute value; right: complex phase) for a source located at $y=1.1$ diffractively lensed by a lens with mass of $3\times10^3M_{\odot}$ assuming the Singular Isothermal Spherical (SIS) profile. 
}
\label{fig:fw}
\end{figure}

To further corroborate this point, it is convenient to define the amplification factor
\begin{equation}
F(f)=\frac{\Tilde{\phi}^{\rm L}(f)}{\Tilde{\phi}(f)},
\end{equation}
where $\Tilde{\phi}^{\rm L}(f)$ and $\Tilde{\phi}(f)$ are the lensed and unlensed GW waveforms. Adopting the thin-lens approximation, the amplification factor at the observer is
\begin{equation}
F(f)=\frac{D_{\rm s}\xi_0^2(1+z_{\rm l})}{D_{\rm l}D_{\rm ls}}\frac{f}{i}\iint d^2\boldsymbol{x}\exp{[j2\pi f t_{\rm d}(\boldsymbol{x,y})]},
\label{eq:af}
\end{equation}
where $\boldsymbol{x}=\boldsymbol{\xi}/\xi_0$, $\boldsymbol{y}=\boldsymbol{\eta}D_{\rm l}/\xi_0 D_{\rm s}$, and $\xi_0$ is the arbitrary normalization constant of the length. The time $t_{\rm d}(\boldsymbol{x,y})$ is the arrival time at the observer from the source, 
\begin{equation}
t_{\rm d}(\boldsymbol{x,y})=\frac{D_{\rm s}\xi_0^2}{D_{\rm l} D_{\rm ls}}(1+z_{\rm l})\left[\frac{1}{2}|\boldsymbol{x}-\boldsymbol{y}|^2-\psi(\boldsymbol{x})+\phi_{\rm m}(\boldsymbol{y})\right],
\end{equation}
where $\phi_{\rm m}$ is chosen to adjust the minimum value of arrival time to be zero. By the above equations, one may therefore be able to estimate the diffractively lensed GW signal. {For example, Figure~\ref{fig:fw} shows the amplification factor $F(f)$ for a Singular Isothermal Spherical (SIS) lens with total mass of $3\times 10^3M_{\odot}$. As seen from this figure, the amplification factor is strongly frequency dependent.}

Here we should point out that if we impose the geometric limit $f\gg t_{\rm d}^{-1}$ (equivalent to ${\lambda_{\rm GW}}/{\lambda_{\rm E}}\ll 1 $), only the stationary points of $t_{\rm d}$ contribute to the amplification factor $F(f)$, for the fast oscillation feature of the integral. Therefore, only images $\boldsymbol{x}_{j}$ satisfying
\begin{equation}
\left.\frac{\partial t_{\rm d}(\boldsymbol{x,y})}{\partial \boldsymbol{x}}\right|_{\boldsymbol{x}_j}=0
\end{equation}
can contribute to the amplification and phase adjustments. In the geometric regime, this partial differential equation reduces to the standard lens equation (see Eq.~\ref{eq:lens}), which embodies the Fermat's principle. Then the integral on the lens plane is reduced to the sum over these images as
\begin{equation}
F(f)=\sum_{i}|\mu_i|^{1/2}\exp{[j2\pi f t_{{\rm d},i}-j\pi n_{i}]},
\end{equation}
where $n_i=0$, $0.5$, and $1$ when $x_j$ is the minimum (Type I), saddle (Type II), and maximum (Type III) points of $t_{\rm d}(\boldsymbol{x,y})$, respectively, representing the parity of the images. Note that Type-II images acquire an additional phase shift of $\pi/2$ relative to Type-I images, as encoded by the Morse phase term $n_i=0.5$. Such a phase shift may introduce characteristic distortions in the observed waveform, especially for signals with more complex morphology, such as higher harmonics, precession, or eccentricity \citep[e.g.,][]{PhysRevD.103.064047,PhysRevD.103.104055,Vijaykumar:2022dlp}. This equation reduces to the geometric-optics expression in the short-wavelength limit, implying that geometric-optics is a valid approximation when diffraction can be ignored.

We note here that calculating lensed waveforms in the wave-optics regime via the amplification factor 
$F(f)$ via Equation~\eqref{eq:af} presents two key challenges. First, from a theoretical standpoint, Equation~\eqref{eq:af} is derived using Kirchhoff’s diffraction theory, which inherently assumes that the incident GW is obstructed by the lens and only accounts for scattered wave contribution (see also Fig.~\ref{fig:f2} for a schematic illustration). When considering that GWs propagate past BHs, fully numerical relativity simulations are necessary to obtain more robust lensed GW waveforms that incorporate the contributions from incident GWs, rather than only scattered components \citep[e.g.,][]{2021MNRAS.506.5278H, 2022PhRvD.106l4037H, 2024PhRvL.132a1401Y}. Second, from a numerical computational perspective, calculating $F(f)$ via Equation~\eqref{eq:af} in the wave-optics regime is both challenging and computationally expensive owing to highly oscillatory nature of the diffraction integral.  Efficient computation of this integral is feasible for axis-symmetric lens profiles \citep{guo=2020}. For more general lensing profiles, the calculation of the diffraction integral can be done by numerical codes implementing fast Fourier transformation, such as \texttt{FIONA}\footnote{https://github.com/ninoephremidze/FIONA} \citep{Ephremidze:2026era} and \texttt{GLoW} \citep{PhysRevD.111.103539}\footnote{https://github.com/miguelzuma/GLoW\_public}.

\section{Source and Lens Populations}
\label{sec:population}

Detailed modeling of GW source and lens populations is essential for the statistical analysis of lensed GW events, as it generates large, reliable mock catalogs for further study. This section introduces some simple models to construct these populations for various lensing systems. 

\subsection{Astrophysical GW Sources}

According to general relativity, GWs can be emitted from celestial systems with a time-varying quadruple moment. Among these, the most interesting astrophysical sources are the inspirals and mergers of binary stellar compact objects and SMBBHs/MBBHs\footnote{In this paper, SMBBHs denote those supermassive ones with total masses roughly in the range $\sim 10^6-10^{10}M_\odot$ and MBBHs denote those with masses roughly $\sim 10^4-10^6M_\odot$.}, which are primary targets of current and future GW observatories, including ground-based detectors, space-based missions, and pulsar timing arrays (PTAs). Observations of these sources can be used to reveal many important astrophysical processes involving in their formation and evolution, such as mass transfer and common envelope evolution of binary stars, and co-evolution of galaxies and their central supermassive black holes (SMBHs). Furthermore, when paired with EM counterparts, they serve as standard sirens, offering independent and precise probes to cosmology that complements existing tools such as the cosmic microwave background (CMB) and Type Ia supernovae. %Beyond these binary sources, the GW emitted by isolated NS and  

In this section, we introduce the cosmic populations of these sources. As currently observations of such sources are still limited, majority of the population studies are focusing on estimating their cosmic distributions, i.e., the cosmic merger rate density of CBCs [$R(m_{1},q;z_{\rm s})$] and the cosmic number density distribution of SMBBHs/MBBHs [$\Phi(M_{\rm tot},q,f_{\rm GW};z_{\rm s})$], via population synthesis models.

\subsubsection{Compact Binary Coalescence}

Binary neutron star (BNSs) mergers, black hole-neutron star (BHNS) mergers and a significant fraction of stellar binary black holes (sBBHs) are believed to originate from the evolution of massive binary stars (hereafter, EMBS channel). One may adopt the cosmic merger rate densities directly obtained from GW observations, however, the uncertainties in these measurements are still large due to small number of detected GW events and lack of high redshift GW events. A widely adopted approach is to combine compact binary population synthesis (CBPS) model and cosmological simulation of galaxy formation and evolution \citep[e.g.,][also see recent reviews by \citealt{Mandeletal2022} and \citealt{2025IJMPD..3442003L}]{2012ApJ...759...52D, 2013ApJ...779...72D, 2015ApJ...806..263D, 2015ApJ...814...58D, 2018MNRAS.479.4391M,2020RAA....20..161H, 2022MNRAS.509.1557C, 2023ApJS..264...45F, 2022MNRAS.509.1557C, 2024ApJ...973..159C, 2025arXiv250814397C}. In principle, the cosmic merger rate density $R^{\rm S}(m_{1},q;z_{\rm s})$ from the EMBS channel may be simply given by \citep[e.g.,][]{2018MNRAS.474.4997C, 2018MNRAS.479.4391M, 2021MNRAS.500.1421Z}
\begin{equation}
R^{\rm S}\left(m_1,q;z_{\rm s}\right) = \int d \tau_{\mathrm{d}}  f_{\rm eff}  R_{\text {birth }}(m_{1},q;{z^{\prime}})  
P_{\tau_{\rm d}}\left(\tau_{\mathrm{d}}\right) , 
\label{eq:Ptaud}
\end{equation}
where $f_{\rm eff}$ denotes the formation efficiency of CBCs, $R_{\text {birth }}(m_{1},q;z^{\prime})$ is the birth rate density of compact binaries with primary mass $m_1$ and mass ratio $q$ at the formation redshift $z^{\prime}$, and $P_{\tau_{\rm d}}$ is the probability distribution of the time delay between the coalescence time of the binary and its formation time ($\tau_{\rm d} =\int^{z_{\rm s}}_{z'} |dt/dz'| dz'$). The birth function $R_{\text {birth }}(m_{1},q;{z^{\prime}})$ can be written as
\begin{equation}
R_{\mathrm{birth}}\left(m_{1},q;{z^{\prime}}\right)  = \iiint d m_{\ast} dq_{\ast} d Z \dot{\psi}\left(Z; {z^{\prime}}\right) \phi\left(m_{\ast}\right) P(q_{\ast}) 
%\times 
 \delta^{(2)}\left(m_{\ast},q_{\ast}| g^{-1}\left(m_{1},q,Z\right)\right), 
\end{equation}
where $\dot{\psi}(Z; {z^{\prime}})$ is the cosmic star formation rate density (SFR) with metallicity $Z$ at the formation redshift $z^{\prime}$, which can be tracked from cosmological hydro-dynamic-simulations for galaxy formation and evolution, 
such as Illustris-TNG \citep{2018MNRAS.475..648P}, EAGLE \citep{2015MNRAS.446..521S}, and GADGET \citep{2021MNRAS.506.2871S} or N-body simulations like Millennium \citep{2005Natur.435..629S} implemented with semi-analytical galaxy formation models \citep[e.g.,][]{2019RAA....19..151J}. The notation $\phi(m_{\ast})$ is the initial mass function (IMF) and $P(q_{\ast})$ is the initial mass ratio of the progenitor binary. The vector function of $g(m_{\ast},q_{\ast},Z)=(m_1,q)$ describes the mapping between the initial zero-age main sequence binaries and the remnant binaries, which can be obtained from the CBPS simulation.
%Note that one may conduct Monte-Carlo Simulations to generate mock catalogue, rather than directly utilize the above expression to obtain merger rate density and its evolution. 
In practice, the evolution of progenitor binaries are simulated by considering several important physical processes, such as mass and angular momentum transfer, common-envelope evolution, natal kick, and mass ejection during SN explosions \citep[e.g.,][]{2002MNRAS.329..897H,2007ApJ...662..504B,2012ApJ...759...52D,2013ApJ...779...72D,2016RAA....16..126W,2016ApJ...819..108B,2022MNRAS.509.1557C, 2022ApJS..258...34R, 2023MNRAS.526.2210S, 2023MNRAS.524..426I, 2024arXiv241215464C, 2025arXiv250602316M, 2025arXiv250814397C,2026RAA....26c2001W}. Then the cosmic merger rate density of CBCs $R\left(m_1,q;z_{\rm s}\right)$ can be estimated by Equation~\eqref{eq:Ptaud} with the time delay $\tau_{\rm d}$ determined in the simulation.

The gravitational interactions of (binary) compact stars in dense stellar systems may also lead to the formation of sBBHs (the dynamical channel) and their subsequent mergers due to GW decay, which may contribute significantly to the sBBH mergers, especially for those with large masses \citep[e.g., ][]{1993Natur.364..423S, 2000ApJ...528L..17P, 2016ApJ...824L...8R,1910AN....183..345V,1962AJ.....67..591K,1962P&SS....9..719L,2016ARA&A..54..441N}. Such dense stellar systems include globular clusters, open clusters and nuclear clusters. 
%The main dynamical processes involved in this channel are binary-binary/single scattering, binary/binary-single exchange, and the formation of hierarchical triple systems \citep[e.g., von Zeipel-Kozai-Lidov mechanism][]{2016ApJ...824L...8R,1910AN....183..345V,1962AJ.....67..591K,1962P&SS....9..719L,2016ARA&A..54..441N}. 
%In the latter one, the von Zeipel-Kozai-Lidov mechanism \citep{1910AN....183..345V,1962AJ.....67..591K,1962P&SS....9..719L} plays an important role, in which the orbit of the inner binary systems undergo periodic oscillations in inclination and eccentricity, due to the perturbative influence of the outer tertiary \citep{2016ARA&A..54..441N}. This effect can facilitate the merger of the inner binary and therefore enhance the merger rate density of sBBHs produced via this channel. 
In principle, the sBBH merger rate density due to dynamical channel can be estimated by full dynamical simulations for sBBH formation in globular clusters, which is however computationally expensive. 
%and it is not possible be done over all parameter spaces. 
An economic way to estimate the merger rate density is to combine the results of sBBH formation given by numerical simulations with simple models for the formation and evolution of globular clusters and other systems over cosmic time. For example, the merger rate density from the dynamical channel for globular clusters may be roughly estimated as \citep[e.g.,][]{2021MNRAS.500.1421Z}:
\begin{equation}
{R^{\rm D}_{\bullet\bullet}}(m_1,q;z_{\rm s})= \left.\iiint \frac{d\dot{M}_{\mathrm{G}}}{d \log M_{\mathrm{H}}}\right|_{z(\tau)} \frac{1}{\left\langle M_{\mathrm{G}}\right\rangle} P\left(M_{\mathrm{G}}\right) R\left(r_{\mathrm{v}}, M_{\mathrm{G}}, \tau-t(z_{\rm s})\right) P(m_1)P(q) d M_{\mathrm{H}} d M_{\mathrm{G}} d \tau,
\label{eq:dyn}
\end{equation}
where $\frac{d\dot{M}_{\mathrm{G}}}{d \log M_{\mathrm{H}}}$ is the comoving SFR in globular clusters per galaxies of a given halo mass $M_{\mathrm{H}}$ at given redshift $z(\tau)$ (or a given formation time $\tau$ ). The function $P(M_{\rm G})$ is the cluster initial mass function, ${\left\langle M_{\mathrm{G}}\right\rangle}$ is the mean initial mass of a globular cluster, while $R(r_{\rm v}, M_{\rm G}, t)$ is the merger rate of sBBHs in a globular cluster with initial virial radius $r_{\rm v}$ and mass $M_{\rm G}$ at time $t(z_{\rm s})$. For simplicity, the specific form for $\frac{d \dot{M}_{\mathrm{G}}}{d \log M_{\mathrm{H}}}$ and $R(r_{\rm v}, M_{\rm G})$ can be  assumed as the explicit function given in \citet{2018ApJ...866L...5R}.
%Among these mock clusters there are  $50\%$ form with $r_{\mathrm{v}} = 1$\,pc and the rest $50\%$ form with $r_{\mathrm{v}} = 2$\,pc.  Then the total rate is the summation of those from the two $r_{\rm v}$ cases. 
Here we also comment that the dynamical channel of BNS mergers \citep[][]{2019MNRAS.488...47F} may be quite inefficient, since NSs are not heavy enough to mostly sink into the centers of globular clusters where the dynamical interactions are most effective \citep{2016ApJ...819..108B}. Notably, we did not take into account the case of nuclear clusters, which may also be promising environments for BH-BH dynamical formation due to their deeper potential \citep[e.g.,][]{2009MNRAS.395.2127O,2019ApJ...881...20R}. 

The sBBH mergers may be also produced in massive accretion disks in Active-Galactic-Nucleus (AGN) around SMBHs (denoted as AGN channel), which was proposed for the origin of some LVK events, such as GW190521, with extremely large masses that cannot be produced directly from stars due to pair instability supernovae explosion \citep[e.g., ][]{2012MNRAS.425..460M, 2017ApJ...835..165B, 2017MNRAS.464..946S, 2019ApJ...877...87Z, 2021ApJ...916L..17W,2022Natur.603..237S, 2022arXiv220407282G, 2024ApJ...961..206C, 2025PhRvD.111h3033M}. A large number of massive stars can be produced in the AGN disk due to the gravitational instability, which may evolve to BHs and dynamically interact with their gaseous environment and other BHs to form sBBH mergers. Recently, it attracts a lot of interests in this channel as it not only can explain the formation of GW190521-like massive sBBHs but also lead to rich EM signals. However, estimating the merger rate density of sBBHs via this channel is difficult due to their complex environments. However, it is plausible to assume that the merger rate density $R(z_{\rm s})$ via the AGN-MBH channel at redshift $z_{\rm s}$ is proportional to the total rate of accretion of mass onto MBHs at that redshift \citep{2022ApJ...940...17C}, 
\begin{equation}
\label{eq:qso}
\boldsymbol{R_{\bullet\bullet}^{\rm A}}(z_{\rm s})\propto \dot{\rho}_{\bullet}^{\mathrm{QSO}}(z_{\rm s}) 
= \int_{0}^{\infty} \frac{(1-\epsilon) L_{\mathrm{bol}}}{\epsilon c^{2}} \Psi(L_{\rm bol}, z) \mathrm{d}L_{\rm bol},
\end{equation}
where $\Psi(L_{\rm bol}, z)$ is the bolometric luminosity function of QSOs, $\epsilon$ is the radiative efficiency of the accretion processes and assumed to be the canonical value $0.1$ \citep[e.g.,][]{2002MNRAS.335..965Y, 2004ApJ...602..603Y, 2004MNRAS.351..169M, 2007ApJ...654..731H, 2008ApJ...689..732Y, 2008MNRAS.383..277K, 2009ApJ...690...20S, 2022MNRAS.509.3488I}. 
%However, the mass and mass ratio distributions for sBBHs produced by the AGN channel is not clear, and one may have to make some simple assumptions about these distributions, which are required for further estimations of their detectability.

Note here that there are many complexities within the above procedures, which may introduce large uncertainties. 
%in estimating the cosmic merger rate density, which may introduce large uncertainties. 
%For example, different cosmological simulation may predict somewhat different star formation and metallicity enrichment histories and therefore result in different absolute $R^{\rm S}(z_{\rm s})$. 
%Different channels produce different merger rate density of sBBHs and their contribution fractions to the total merger rate density are different. %Basically, the total merger rate density and its evolution determined by the local merger rate density and the shape of the evolution function. 
Nevertheless, LVK observations have already put strong constraints on the local merger rate densities, i.e., $19_{-5}^{+7} \rm Gpc^{-3} yr^{-1}$ for sBBHs, $89_{-67}^{+159} \rm Gpc^{-3} yr^{-1}$ for BNSs, and $23_{-13}^{+20} \rm Gpc^{-3} yr^{-1}$ for BHNSs, respectively \citep{2025arXiv250818083T}. Probably most BNS and BHNS mergers are produced by the EMBS channel, therefore, the local merger rate densities of BNSs and BHNSs may be rescaled by those obtained by LVK observations. 
%If assuming most sBBHs mergers are also produced by the EMBS channel, one could use the observationally determined local merger rate to calibrate the merger rate density resulting from this channel. 
For the dynamical channel, however, the value of $R^{D}_{\bullet\bullet}$ is quite difficult to estimate. Nevertheless, \citet{2018ApJ...866L...5R} obtained $R^{D}_{\bullet\bullet} \sim 14_{-10}^{+4} \rm Gpc^{-3} yr^{-1}$ using a simple model for globular clusters, which may be adopted as a rough calibration. For open clusters and nuclear clusters, it is also possible to estimate the merger rate density with similar implementation in Equation~\eqref{eq:dyn}, but with substantial uncertainties.
%, both due to the poor understanding of their cosmological evolution and the distributions of their intrinsic properties. 
For the AGN channel, it is also challenging to estimate the merger rate density accurately due to various complications, though Equation~\eqref{eq:qso} offers a simplified evolutionary approximation. Among these complexities, the most significant source of uncertainty lies in the local merger rate density estimates, which span an exceptionally wide range—from $0.02$ to $60$\,Gpc$^{-3}$\,yr$^{-1}$ \citep[e.g.,][]{2017MNRAS.464..946S,2018ApJ...866...66M,2019MNRAS.488...47F,2020ApJ...898...25T,2024MNRAS.529..883I}.

\begin{figure}
\centering
\includegraphics[width=0.8\columnwidth]{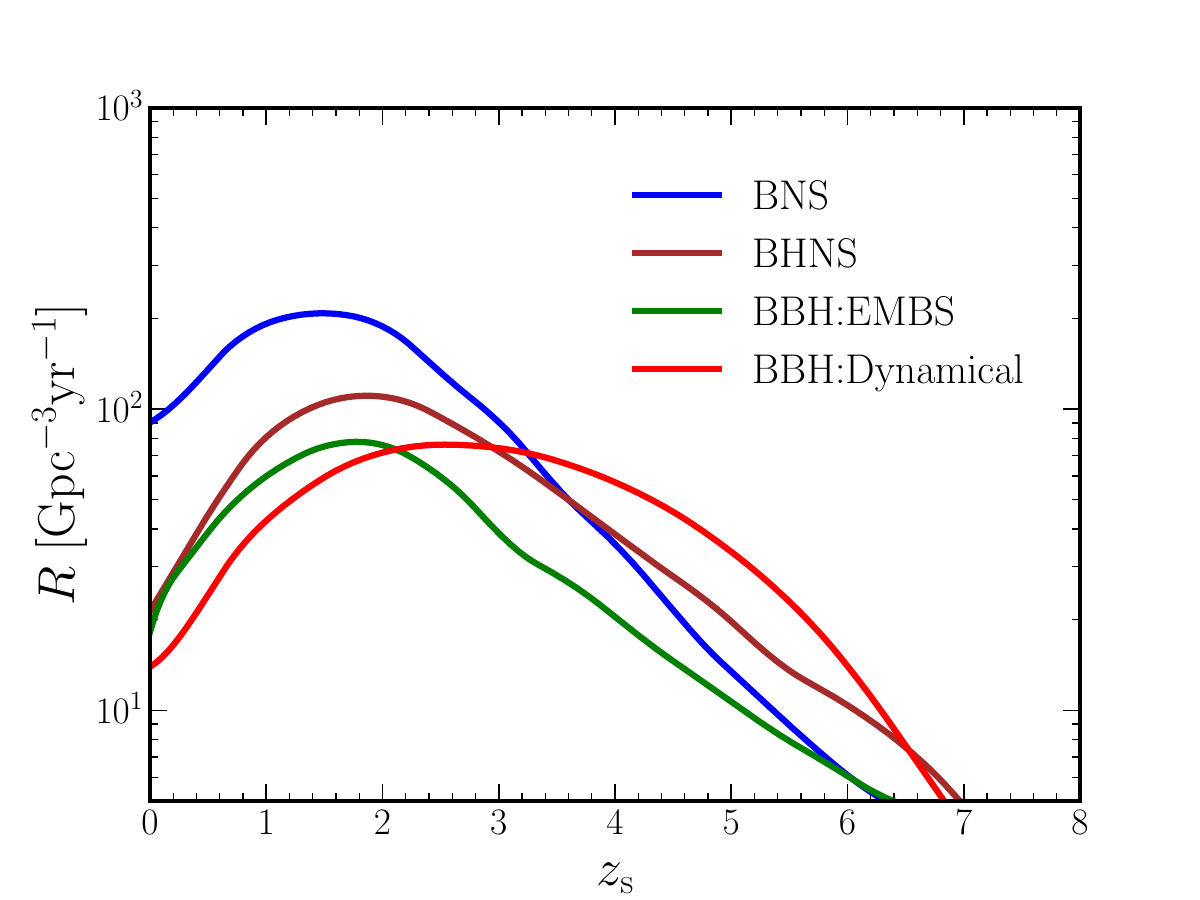}
\caption{
The cosmic merger rate density evolution $R(z_{\rm s})$ of compact binary coalescence, including BNS (blue), BHNS (brown), and BBH produced via EMBS (green) and dynamical (red) channel \citep[e.g.,][]{2020A&A...636A.104B,2021MNRAS.500.1421Z}. The results are rescaled by the latest constraint on the local merger rate density from GWTC-4.0 \citep{2025arXiv250818083T}, i.e., $R^{\rm E}\sim 19 \rm Gpc^{-3} yr^{-1}$ for sBBHs produced via EMBS channel, $89 \rm Gpc^{-3} yr^{-1}$ for BNSs, and $23 \rm Gpc^{-3} yr^{-1}$ for BHNSs. The merger rate density for BBH produced via dynamical channel are rescaled by $R^{\rm D}\sim 14 \rm Gpc^{-3} yr^{-1}$ according to \citet{2018ApJ...866L...5R}. 
}
\label{fig:merger_rate}
\end{figure}

\subsubsection{Supermassive Binary Black Holes}
\label{sec:MBBH}

SMBBHs/MBBHs are natural products of galactic mergers as SMBHs are ubiquitous in galactic centers \citep[e.g.,][]{Kormendy=2013}. The population properties and merger rate of SMBBHs/MBBHs are set by the cosmic history of galaxy mergers, the masses (and spins) of the central SMBHs in the progenitor galaxies, and the subsequent orbital evolution of SMBBH pairs during the merger and within merger remnant.
To date, due to the absence of robust detections of individual SMBBHs from either EM \citep[e.g.,][]{runnoe2017,guo2019,chen2023a,huijse2025} or PTA \citep[e.g.,][]{NANO23cgw15,EPTA24cgwdr2} observations, 
%Survival SMBBHs with period on the order of $1-10$\,yr radiate nanohertz GWs that may be detectable by pulsar timing arrays (PTAs), while merging MBBHs with mass on the order of $10^4-10^7M_\odot$ emit millihertz GWs that may be detectable by future space-based GW detectors. Although recent PTA observations provide evidence for a nanohertz gravitational-wave background, likely produced by inspiralling SMBBHs across cosmic time, no individual SMBBHs has yet been resolved \citep[e.g.,][]{NANO23cgw15,EPTA24cgwdr2}. EM observations have identified several hundreds SMBBH candidates through spectral signatures \citep[e.g.,][]{runnoe2017,guo2019} and periodic variability \citep[e.g.,][]{chen2023a,huijse2025}, but robust confirmation remains difficult, consequently,
the intrinsic population properties and merger rate of SMBBHs cannot yet be inferred from observations directly \citep[see][]{wang2020a,2025CQGra..42k3001L}. Therefore, estimates of SMBBH population properties and merger rates must rely on theoretical modelling, either semi-analytical approaches that combine observational constraints on the model ingredients \citep[e.g.,][]{Sesana13,2020ApJ...897...86C} or fully hydrodynamic cosmological simulations of galaxy formation and evolution \citep[e.g., ][]{2015MNRAS.450.1937C, 2016A&C....15...72M,2018MNRAS.475..676S, 2018MNRAS.475..648P}. The former offers great flexibility for exploring large parameter spaces, while the latter captures more detailed galaxy formation physics and the complex interactions between SMBHs/SMBBHs with their hosts. Below we introduce the framework for a semi-analytic SMBBH model based on \citet{2020ApJ...897...86C} with detailed consideration of the SMBBH orbital evolution, while for the SMBBH model through hydrodynamic cosmological simulations one may find in \citet{Kelley17smbbh} and \citet{Katz20}. 

The cosmic distribution $\Phi(z_{\rm s}, M_{\rm tot},q,f_{\rm GW})$, defined as the comoving number density of SMBBHs at redshift $z_{\rm s}$ per unit total mass $M_{\rm tot}$, per mass ratio $q$, per GW frequency $f_{\rm GW}$, can be estimated straightforwardly by as \citep[c.f.,][]{2020ApJ...897...86C},
\begin{equation}
\begin{aligned}
\Phi_{\mathrm{BBH}} \left(M_{\mathrm{BH}}, q_{\mathrm{BH}}, f_{\mathrm GW}, z_{\rm s}\right) 
\simeq & \int_0^t d t^{\prime} \int d M_* \int d q_* n_*\left(M_*, z^{\prime}\right) \mathcal{R}_*\left(q_*, z^{\prime} | M_*\right) \times \\
&  p_{\mathrm{BH}}\left(M_{\mathrm{BH}}, q_{\mathrm{BH}} | M_*, q_*, z^{\prime}\right) 
%\\ & 
%
%\times 
p_f(f_{\mathrm GW}, t-t^{\prime} | M_*, q_*, M_{\mathrm{BH}}, q_{\mathrm{BH}}, z^{\prime}). 
%\\&\times 
%P_{\text {intact }}\left(z_{\rm s}, z^{\prime} | M_*\right).
%
\end{aligned}
\label{eq:SMBBH_ND}
\end{equation}
Here $n_*(M_*,z')$ is the stellar mass function of the spheroidal components of the merger remnants (i.e., bulge) with mass $M_*$ at redshift $z'$ and $\mathcal{R}_*(q_*,z'|M_*)$ represents the averaged number of mergers experienced by spheroids with masses $M_*$ at redshift $z'$ per unit mass ratio per unit time. These terms related with the progenitor bulges can be derived from quantities of their host galaxies, i.e., 
\begin{equation}
\begin{aligned}
n_* &(M_*, z^{\prime}) \mathcal{R}_*(q_*, z^{\prime} | M_*) = \\
& \iint d M_{\mathrm{gal}} d q_{\mathrm{gal}}  n_{\mathrm{gal}}(M_{\mathrm{gal}}, z^{\prime}) \mathcal{R}_{\mathrm{gal}}(q_{\mathrm{gal}}, z^{\prime} | M_{\mathrm{gal}})
%\\
%\quad & \times 
p_*(M_*, q_* | M_{\mathrm{gal}}, q_{\mathrm{gal}}, z^{\prime}),
\end{aligned}
\end{equation} 
where $n_{\rm gal}(M_{\rm gal},z')$ and $\mathcal{R}_{\rm gal}(q_{\rm gal},z'|M_{\rm gal})$ are the galaxy stellar mass function and the merger rate by galaxies with masses $M_{\rm gal}$ at redshift $z'$ per unit mass ratio $q_{\rm gal}$, respectively; $p_*(M_*,q_*|M_{\rm gal},q_{\rm gal},z')$ is the probability distribution of the bulge total mass and mass ratio at redshift $z'$ given the host galaxy total mass and mass ratio. All these galaxy quantities are defined in a similar way as the bulges. The above terms can be estimated with the help of the cosmic merger tree produced by cosmological hydrodynamical simulations \citep{2018MNRAS.475..648P,2015MNRAS.446..521S}.  

The functions $p_{\rm BH}(M_{\rm BH},q_{\rm BH}|M_*,q_*,z')$ and $p_f(f_{\rm GW},\tau| M_*, q_*, M_{\rm BH}, q_{\rm BH}, z')$ in Equation~\eqref{eq:SMBBH_ND} are two probability distribution functions to take into account the BH-galaxy relationship, and the SMBBH orbit evolution. The first one is defined so that $p_{\rm BH}(M_{\rm BH},q_{\rm BH}|M_*,q_*,z') dM_{\rm BH} dq_{\rm BH}$ represents the probability that a host merger at redshift $z'$ characterized by $(M_*,q_*)$ leads to a SMBBH merger characterized by $(M_{\rm BH}, q_{\rm BH})$ with $M_{\rm BH}$ representing the total mass of the SMBBHs, which is inferred by the MBH--host galaxy scaling relation \citep[e.g.,][]{2023OJAp....6E..27G};
The second one is defined so that $p_f(f_{\rm GW},\tau|M_*,q_*, M_{\rm BH}, q_{\rm BH}, z')df_{\rm GW}$ represents the probability that a host merger characterized by parameters $(M_*,q_*,M_{\rm BH},q_{\rm BH})$ at redshift $z'$ leads to an SMBBH emitting GW at frequency range $f_{\rm GW}\rightarrow f_{\rm GW}+df_{\rm GW}$ at a time $\tau$ after the host merger. This term encodes the key dynamical processes that SMBBHs experience as they evolve from galaxy pairing to final coalescence. Briefly speaking, the evolution tracks of an SMBBH inside a galaxy merger remnant can be divided into several stages, including (1) dynamical friction stage, when the orbits shrinks mainly due to distant star field and gas particle scattering \citep[e.g.,][]{2001ApJ...563...34M, 2002MNRAS.331..935Y}; (2) binary stage with orbital decay governed by interactions with stars and gas material in its environment, first jointly by dynamical friction and three-body scattering of stars passing through the SMBBH (non-hard binary) \citep[e.g., ][]{1996NewA....1...35Q, 2002MNRAS.331..935Y}, then dominated by the three-body star scattering (hard binary) \citep[e.g.,][]{1996NewA....1...35Q,2002MNRAS.331..935Y} and/or by viscous torque from an accretion disk around the binary if the merger occurs in a gas-rich environment; (3) GW radiation dominant stage, with the orbit decay rapidly driving by the GW radiation toward the final coalescence (responsible for nanohertz GW insprialing SMBBHs). The SMBBH orbital evolution have been studied extensively and numerous mechanisms have been proposed to accelerate orbit decay and drive coalescence on timescales shorter than the cosmic (Hubble) time \citep[e.g.,][]{2025CQGra..42k3001L}. Assuming the delay between galaxy merger and SMBBH merger is substantially smaller than the Hubble time (e.g., $\sim10^9$\,yr or shorter), which may be true for most SMBBH major mergers in small galaxies, then the merger rate of SMBBHs can be estimated as
\begin{equation}
\dot{\Phi}_{\mathrm{BBH}} (M_{\rm BH}, q_{\rm BH}; z) \sim \int d M_* \int d q_* n_*\left(M_*, z\right) \mathcal{R}_*\left(q_*, z | M_*\right) 
 p_{\mathrm{BH}}\left(M_{\mathrm{BH}}, q_{\mathrm{BH}} | M_*, q_*, z^{\prime}\right). 
\end{equation}
%In the above Equations, we neglect the possibility of triple interactions in which the central SMBBH in a merged galaxy subsequently undergoes a major merger with a third SMBH delivered by a later galaxy merger. In principle, this can be roughly estimated \citep[e.g.,][]{2020ApJ...897...86C}.
More details on the above descriptions and the generation of mock SMBBH samples via Monte-Carlo simulation can be seen in \citet[][see aslo \citealt{2023ApJ...955..132C}]{2020ApJ...897...86C}. 
%We also note that in the above formalism, the BHs are assumed to have nearly circular orbits and thus therefore ignore their eccentricity evolution. However, there are still debates on the eccentricity of SMBBHs, arguing that the scattering of ambient stars and interaction with circumbinary disks may also lead to some eccentric SMBBHs. 

The population synthesis model described above relies on the galaxy stellar mass function and merger rate with constraints from observations, which is effectively obtained only up to redshift $z\sim 5-6$ because of the paucity of high‑redshift observations. Extrapolating this to low-mass end of SMBBHs/MBBHs (e.g., $\lesssim 10^5-10^6M_\odot$) therefore may have substantial uncertainty. When considering merger rate of SMBBHs/MBBHs at low-mass end that may be detected by space-based GW detectors, more comprehensive approaches are required, which must follow galaxy assembly to much earlier times ($z\gtrsim 10-20$) and explicitly implement SMBH seeding; however seeding physics remains poorly understood and different seeding prescriptions produce large variations in predicted merger rates. For example,  \citet{Barausse2012} developed a semi-analytic model (SAM), based on the extended Press-Schechter (EPS) merger trees, to track mergers of host halos and galaxies across cosmic time and generate SMBBH catalogs for both light-seed ($\sim 10-100M_\odot$) and heavy-seed ($\sim10^4-10^6M_\odot$) scenarios.  Though, such SAM has since been refined to include more realistic physics, including accretion \citep{Sesana2014}, co-evolution with nuclear star cluster (NSC) \citep{Antonini2015a,Antonini2015b}, the time-delay between galaxy and SMBH mergers \citep{Klein2016,Bonetti2018,Bonetti2019}. kpc-scale wandering, and supernova feedback \citep{Barausse2020}, the SMBBH merger rate predictions remain highly model dependent and that progress requires both deeper high‑redshift observations and improved theoretical modeling of SMBH seeding and early growth. In addition, we also note other works in the literature  \citep[e.g.,][]{Jiang2019,Wang2019, Salcido2016, Izquierdo2019,Izquierdo2020, Izquierdo2022} that directly adopted cosmological numerical simulation results to trace merger history of dark matter halos and galaxies, with implementations of SMBH seeding models and galaxy formation processes, to estimate SMBBH merger rates.

\subsection{Lens Population}

The propagation of GW can be deflected by any massive intervening objects/systems as discussed above in Section~\ref{sec:basic}. In classical gravitational lensing studies of light propagation, the lensing effect is typically classified into three categories according to the nature of the lensing mass \citep{Schneider=2006}: (1) strong lensing, caused by systems like galaxies and galaxy clusters; (2) microlensing, induced by objects such as individual stars, BHs, or planets; and (3) weak lensing, resulting from the cumulative influence of large scale structures in the universe. Similarly, GW lensing includes strong lensing by galaxies and galaxy clusters and microlensing by stars, black holes, or planets \citep[e.g.,][]{PhysRevD.109.024064,2021SCPMA..6420462Z,2025arXiv250821262S}; for certain combinations of source frequency and lens Einstein radius, wave‑optics effects must be taken into account. However, weak lensing is sparingly studied for GWs and mentioned in the context of the stochastic background or as a noise term in distance measurement \citep[e.g.,][]{2024PhRvD.110b3502M,2024MNRAS.533...36C}. A newly recognized intermediate regime between wave- and geometric-optics — millilensing — features limited diffraction yet overlapping lensed GW signals whenever the inter‑image time delays are shorter than the waveform duration, yielding a frequency‑dependent interference (beating) pattern \citep[e.g.,][]{2023MNRAS.525.4149L,2024PhRvD.110l3008L,2025PhRvD.112f4043L}. We refer readers to the above cited works for detailed treatments and limit our scope to lenses composed of galaxies, galaxy clusters, stars, and dark matter minihalos in both geometric and wave-optics regimes.

\subsubsection{Galaxies}
\label{sec:galaxy}

Galaxies as the lens are the most common components in the gravitational lensing systems in universe, for their large mass and abundances. A large amount of effort have been put into both the theoretical and observational studies of galaxy-galaxy, Quasar-galaxy, and supernova-galaxy lensing systems \citep[e.g.,][]{2015ApJ...803...71S,2017ApJ...851...48S,2021AAS...23755501S,2022ApJ...924...49C,2022RAA....22k5019D,2023RAA....23c5001Z, 2024SSRv..220...13S, 2024AJ....167..264L, 2024MNRAS.533.1960C, 2024arXiv241118694M, 2024A&A...681A..68E,2025MNRAS.540.3121C}. These systems can be used as probes to constrain cosmological parameters \citep[e.g.,][such as the Hubble constant $H_0$ and the amplitude of the density fluctuation $\sigma_8$]{2020MNRAS.498.1420W}, as well as the density profile of galaxy dark matter halos \citep[e.g.,][]{2012ApJ...744..159L, 2009ApJ...704.1274W}.

Strong lensing may be dominated by elliptical (early-type) galaxies (spiral galaxies may also contribute mildly`, see \citet{2014JCAP...10..080B} ), for their larger mass $\sim 10^{8}-10^{12} M_\odot$ and compact density profile (with Sersic index $\sim4$) \citep[e.g.,][]{Oguri_2010, 2017MNRAS.465.4895W}. The lens galaxies are therefore typically modelled by the Singular-Isothermal-Ellipsoidal (SIE) profile\footnote{It is also frequently simply modelled as the Singular-Isothermal-Sphere (SIS) for rough estimations.} \citep{1994A&A...284..285K}, of which the convergence in the Cartesian coordinates can be written as
\begin{equation}
\kappa(x,y)=\frac{1}{2}\frac{\lambda(q_l)\sqrt{q_l}}{\sqrt{x^2+q_l^2y^2}},
\end{equation}
where $q_l$ is the projected minor-to-major axis ratio and $\lambda (q_l)$ is the dynamical normalization, dependent on the three-dimensional shape of the lensing galaxies. 
Then, the first derivative of lens potential (Eq.~\ref{eq:potential}) for SIE profile can be derived as
\begin{equation}
\phi_x=\frac{b_I(q_l)}{e}\text{arctanh}\left[ \frac{ex}{\Delta}\right], \ \ 
\phi_y=\frac{b_I(q_l)}{e}\arctan\left[\frac{ey}{\Delta}\right],
\end{equation}
where $b_I(q_l)=\lambda(q_l)\sqrt{q_l}$, $\Delta=\sqrt{x^2+q_l^2y^2}$, and $e=\sqrt{1-q_l^2}$ is the ellipticity of the lensing galaxy. In this case, the characteristic length $\xi_0$ is defined as
\begin{equation}
\xi_0=4\pi\frac{\sigma_{\rm v}^2}{c^2}\frac{D_{\rm l}D_{\rm ls}}{D_{\rm s}},
\end{equation}
which is directly related to the velocity dispersion $\sigma_{\rm v}$ of the lens galaxy.

For population studies of the lensing systems, the velocity dispersion function (VDF) $dn(\sigma_{\rm v},z_{\rm l})/dz_{\rm l}$ of lens galaxies is required, which may be given by observations with a fitting form as \citep[e.g.,][]{2007ApJ...658..884C, Piorkowska:2013eww}:
\begin{equation}
\frac{dn(\sigma_{\rm v},z_{l})}{d\ln \sigma_{\rm v}}=n_z \frac{\beta_g}{\Gamma(\alpha_g/\beta_g)}
%
%\frac{1}{\sigma_v} 
\left(\frac{\sigma_{\rm v}}{\sigma_z}\right)^{\alpha_g}\exp{\left[-\left(\frac{\sigma_{\rm v}}{\sigma_z}\right)^{\beta_g}\right]},
\label{eq:vdf}
\end{equation}
and
\begin{equation}
{n_z = n_{0}(1+z)^{\kappa_{n}};\quad \sigma_z = \sigma_{\rm v0}(1+z)^{\kappa_{\rm v}}}.
\label{eq:nzsigmaz}
\end{equation}
Here $\sigma_{\rm v0}$ is the characteristic velocity dispersion, $\alpha$ is the power-law index at the low-velocity end, $\beta$ is the high-velocity exponential truncated index, and $\Gamma(\alpha_g/\beta_g)$ is the Gamma function. The reference parameters in this formula are chosen to be the canonical values, i.e.,  $(n_0, \sigma_{\rm v0}, \alpha_g, \beta_g) = (0.008h^{3} {\rm Mpc}^{-3}, 161{\rm km \, s^{-1}}, 2.32, 2.67)$ given by SDSS observations \citep[e.g.,][]{2007ApJ...658..884C}. The evolution parameters $\kappa_n$ and $\kappa_{\rm v}$ are given the fitting results from \citet{2021MNRAS.503.1319G}, i.e., $\kappa_n = -1.18$ and $\kappa_{\rm v}=0.18$. The distribution of axis ratio $P(q)$ is modelled by a truncated Gaussian distribution in the range of $[0.2,1]$, with a mean of $0.7$ and a standard deviation of $0.16$, to match the observations of early-type galaxies \citep{2003ApJ...594..225S}. According to the above Equations~\eqref{eq:vdf} and \eqref{eq:nzsigmaz}, mock lens elliptical galaxies can be generated by Monte-Carlo method and therefore the properties of the lensed GW events can be estimated. These properties include the magnification factor and time-delay in the geometric-optics regime and the frequency dependent amplification factor in the wave-optics regime. 

In the geometric-optics regime, the number of images due to strong lensing may depend on the lens density profile \citep[e.g.,][]{2015A&A...580A..79T,2021ApJ...921..154W}. When assuming the SIE profile, the number of images with finite magnification factors are $2$ or $4$, and the fraction of $4$-image cases is much smaller comparing with that of $2$-image cases. However, adopting the more realistic lens models, such as Non-Singular Isothermal Ellipsoidal (NSIE) or Double Elliptic Power-Law (DEPL) - motivated by the existence of core-like structure in early-type galaxies \citep{1996AJ....111.1889B, 1997ASPC..116..113L}- can produce additional images (including a central image) that do not appear in contrast to the simpler SIE model. The central image, arising from the core-like structure of the lens galaxy, has traditionally received limited attention for several reasons. First, it is normally much fainter than the other images due to its low (de)magnification factor, typically around $\mu \sim 0.01$. Second, its location is close to the optical axis of the lens system, making it susceptible to strong contamination from the lens galaxy's own emission. Nonetheless, incorporating a core-like structure into the lens galaxy is a common practice when modeling member galaxies in cluster lensing systems \citep[e.g.,][]{2024MNRAS.531.1179X}. Notably, two radio-detected lensed Quasars, with three \citep{2004Natur.427..613W} and five images \citep{2005PASJ...57L...7I}, respectively, show central images that are plausibly attributable to core-like mass distribution in their lensing galaxies. When considering lensed GW events, the above two difficulties in detecting the central image, if any, may be naturally overcome. First, the S/N of the central image may be only about ten times fainter than other images as its signal $\propto \sqrt{\mu}$, which may be still detectable by the third generation GW detector despite that it is relatively weak comparing with other images. Second, the GW signal can get through the lens galaxy without any contamination though it may be affected slightly by the micro-lensing effect. It has been shown that the detection of the central image is probable \citep{zhiwei_na}.

\subsubsection{Galaxy clusters}

Galaxy clusters, with total masses of $10^{13}- 10^{15}M_{\odot}$ , are the most massive self-gravitationally bound structures in the Universe, which can also act as lenses to produce even stronger lensing effect than galaxies do \citep[e.g.,][]{2002ApJ...568L..75H, 2012ApJS..199...25P,2015Sci...347.1123K,2019ApJ...878..122L,2025arXiv251106928X}. Gravitational lensing by galaxy clusters are currently a powerful tool for studying various aspects of astrophysics and cosmology, for instance, probing the highest redshift faint galaxies and providing insights into the origin and evolution of galaxies \citep[e.g., ][]{2015ApJ...812..114T, 2022Natur.603..815W}, constraining the nature of dark matter \citep[e.g.,][]{2020ApJ...898...87G, 2022A&A...657A..83C} and provide inference to cosmological parameters \citep[e.g.,][]{2025MNRAS.544..708X}. Though their abundance is much smaller than galaxies, but due to their comparatively larger masses, the cluster lensing of GW events may be as important as galaxy lensing \citep[e.g.,][]{2023MNRAS.520..702S, 2024ApJ...977...64C}. 

Modeling galaxy clusters is challenging because of their complex structures-multiple clumpy dark matter halos and numerous dynamically perturbed substructures. However, \citet{2024ApJ...977...64C} found that the overall detection rate of the cluster-lensed GW events, whether estimated from mock clusters in the Synthetic Sky Catalog for Dark Energy Science with LSST (cosmoDC2) \citep[e.g.,][]{korytov19,lsst12} or derived using an eccentric Navarro–Frenk–White \citep[eNFW;][]{1997ApJ...490..493N,Wright2000ApJ,Golse2002} dark matter halo plus a brightest central galaxy (BCG) with SIS profile, are in agreement with each other. This consistency arises because small-scale substructures introduce only minor perturbation to the combined lensing potential of the BCG and the dark matter halo. Therefore, adopting the eNFW+SIS profile offers a plausible and efficient approximation for estimating the detectability of cluster-lensed GW events. On the one hand, the density profile of the NFW profile is
\begin{equation}
\rho_{\rm NFW}(r)=\frac{\rho_{\rm NFW}}{\frac{r}{r_{\rm s}}\left(1+\frac{r}{r_{\rm s}}\right)^2},
\label{eq:nfw}
\end{equation}
where $\rho_{\rm NFW}$ is the characteristic density, $r_{\rm s}=r_{\rm vir}/c_{\rm v}$ is the characteristic length, which can be estimated according to the mass $M_\ell$, concentration $c_{\rm v}$, and virial radius $r_{\rm vir}$ of the dark matter halo. Note that the estimate of the optical depth is strongly dependent on the adopted relationship between $c_{\rm v}$ and  $M_\ell$. A simple intuition is that the higher concentration $c_{\rm v}$ with the same $M_\ell$, the larger the cross-section due to relatively stronger deflection of the GW. On the other hand, the central BCG is assumed to follow the SIS density profile as 
\begin{equation}
\rho_{\rm SIS}(r)=\frac{\sigma_{\rm v}^2}{2\pi G r^2},
\end{equation}
where $\sigma_{\rm v}$ is the velocity dispersion of the BCG. The BCG  can be linked to the main dark matter halo by adopting the extended empirical $M_{\ast,\rm BCG}-M_{\ell}$ relationship \citep[for example, see][]{2019A&A...631A.175E}. Then the velocity dispersion $\sigma_{\rm v}$ of the BCG can be estimated via the BCG mass according to observations \citep[for example, see][]{2009MNRAS.394.1978H}. Notably, unlike the case for galaxy lensing, the above joint lens model may only be solved numerically rather than analytically, for their complex math structures.

In the above eNFW+SIS model for cluster lensing, the key parameter is the dark matter halo mass $M_\ell$. The distribution of the lens cluster population is given by the cosmic halo mass function (HMF). For example, a widely adopted general expression is given by \citet{1999MNRAS.308..119S}
\begin{equation}
\frac{dn_{\rm H}}{dM_{\ell}}\propto \frac{\rho_0}{M} \left| \frac{d\sigma}{dM} \right| (1+(a\nu^2)^{-p})\sqrt{a\nu^2} \exp\left(-\frac{a\nu^2}{2}\right),
\label{eq:hmf}
\end{equation}
where $\rho_0$ is the mean mass density of the universe, $\sigma(M)$ is the variance of density fluctuations in the universe on a scale corresponding to mass $M$, and $\nu$ is the  dimensionless parameter defined as $\nu = \delta_{\rm c}/\sigma(M)$, in which $\delta_{\rm c}$ is the critical density threshold for collapse (approximately $\delta_{\rm c} \approx 1.686(1+0.123\log_{10}{}\Omega_{\Lambda})$ depending on the dark energy density $\Omega_{\Lambda}$ \citep{1996ApJ...469..480K}). The fitting parameters $a$ and $p$ are suggested to be $(a,p)=(0.75,0.3)$ by \citet{1999MNRAS.308..119S} and notably when $(a,p)=(1,0)$, it restores the original Press-Schechter function \citep{1974ApJ...187..425P}. We refer the interested readers to Python package \texttt{hmf} to generate HMF in various forms \citep{2013A&C.....3...23M}. The HMF behaves approximately as a power-law over $10^{13}-10^{14} M_\odot$ and declines rapidly above $10^{14} M_\odot$, implying that strong lensing by galaxy clusters is dominated by the low mass end of the cluster population.  

\subsubsection{Stars}

Stars within galaxies can also play a role in gravitational lensing as the intervening object, which corresponds to the microlensing regime \citep[e.g.,][]{1986ApJ...304....1P,1991ApJ...374L..37M,1997ApJ...486..697A,2006Natur.439..437B,2012RAA....12..947M,2012ARA&A..50..411G,2014ApJ...784..100Y,2021MNRAS.508.4869M,2021MNRAS.508.1253S,2025arXiv250821262S}. As for the GW lensing, the condition $\lambda_{\rm GW}/\lambda_{E}\gtrsim 1$ is always satisfied in this case. Therefore, the gravitational lensing of GW by stars should be considered within the wave-optics regime. Due to the small size of stars, we often model them as superposition of $N$ point sources, of which the lensing potential is
\begin{equation}
\phi(x)=\sum_{i}^{N}\frac{M_{\ast,i}}{\langle M_{\ast}\rangle}\ln{(x_i-x)^2},
\end{equation}
where $M_{\ast,i}$ is the mass of the $i$-th star and $\langle m_{\ast}\rangle$ is the average mass of stars. Note that the microlensing by stars in the lenses always occur, therefore the distribution of stars is largely dependent on the host galaxies. The generation of mock star samples in a single galaxy usually involves the following steps \citep[e.g.,][]{2023arXiv230106117S}: (1) For each pixel of the galaxy, sample a total mass of stars, following the surface brightness distribution of the galaxy, for example elliptical Sersic law. (2) The mass distribution of stars in each pixel obeys a initial mass function (IMF). (3) The remnant objects as lens may also be considered, given the relationship between the initial mass to the remnant mass.  

Although constructing the star samples and models is relatively straightforward, evaluating the diffraction-lensing integral becomes extremely challenging for large $N$, giving rise to convergence, precision, and excessive computation-time issues. To overcome these hurdles, \citet{2025SCPMA..6819512S} recently introduced the Trapezoid Approximation-based Adaptive Hierarchical (TAAH) Tree  algorithm. In TAHH algorithm, one first calculates the lensing amplification factor in the time domain, and then applies an inverse Fourier transformation within a proper choosing lens plane boundary. They found that the lensing potentials produced by far-field stars are smooth and can be calculated using interpolation techniques based on sparsely sampled grids. In contrast, the near-field stars exhibit rapid potential variation and requiring finer grids. By adjusting dynamically the grid resolution in the galaxies, it can significantly reduce computation time while still preserving numerical accuracy. 

\subsubsection{Primordial Black holes}

The primordial black holes (PBH) are originated from the gravitational collapse of the overdense regions in the early universe \citep[][]{1967SvA....10..602Z,1971MNRAS.152...75H}, which is a fancy candidate for dark matter. The optical microlensing effect of PBHs has been widely used to constrain the mass of PBHs. In principle, PBHs can also serve as lenses for the GW signals \citep[e.g., ][]{2023JCAP...03..043C, 2023PhRvD.108b3507U}. Here we divide the PBH lensing into two categories. The first one is for bare PBHs, which can be modeled as simple point mass lenses \citep[e.g.,][]{1998PhRvL..80.1138N,Deka:2024ecp}. The second one is for dressed PBHs, considering the minihalo around each PBH caused by accretion of surrounding dark matter  \citep[e.g.,][]{1985ApJS...58...39B,2016AstL...42..347E}. In the latter case, the density profiles of the surrounding minihalos are often parameterized as \footnote{The power law index $-9/4$ is originally obtained by \citet{1985ApJS...58...39B} assuming spherical collapse in Einstein-de Sitter Universe without central BHs, which may not be the case for dressed PBHs. Analogous to the collisionless particles around the BH, the density profile of dark matter halo may be sketched by the Bahcall-Wolf cusp with index of $-7/4$ derived from galactic dynamics \citep{1976ApJ...209..214B} rather than $-9/4$.} \citep[e.g.,][]{2019PhRvD.100b3506A, 2023JCAP...03..043C}:
\begin{equation}
\rho_{\mathrm{}}(r)=
\begin{cases}
\frac{3M_{\rm halo}}{16\pi R_{\rm halo}^3}\left(\frac{r}{R_{\rm halo}}\right)^{-9/4}, & \mathrm{for~}r\leq R_{\rm halo}, \\
0, & \mathrm{for~}r>R_{\rm halo},
\end{cases}
\end{equation}
where $M_{\rm halo}$ is the total mass of the halo and $R_{\rm halo}$ is the cutoff radius of the halo density profile. These two quantities are directly related to the PBH mass and the fraction of dark matter in the forms of PBH, i.e., $m_{\rm PBH}$ and $f_{\rm PBH}$. A simple argument is that $M_{\rm halo}$ is limited by the available amount of particle dark matter in the universe, which implies that $f_{\rm PBH}M_{\rm halo}\leq (1-f_{\rm PBH})m_{\rm PBH}$. The equality is only satisfied in the idealized case when all of the dark matter is contained in PBH minihalos. One may also expect a stronger condition that the minihalo growth is stopped once the PBHs are absorbed into nonlinear dark matter structures, which imposes a limitation such as $1+z_{\rm NL}\sim 5500f_{\rm PBH}$ \citep[e.g.,][]{2019PhRvD.100h3528I}, where $z_{\rm NL}$ is the redshift for the formation of nonlinear dark matter structures. Then, the size of the dressed PBH minihalo $R_{\rm halo}$ can be defined by the radius at which the density of the minihalos matches the ambient dark matter density, i.e., $\rho_{\rm R_{\rm halo}}\sim \rho_{\rm DM}(z_{\rm NL})$. Therefore, we have
\begin{equation}
R_{\rm halo}\sim 14{\rm pc} \left(\frac{m_{\rm PBH}}{M_{\odot}}\right)^{1/3} {\rm min}(1,10^{-3}f_{\rm PBH}^{-4/3}),
\end{equation}
and 
\begin{equation}
M_{\rm halo}\sim (1-f_{\rm PBH})m_{\rm PBH} {\rm min} (50,0.28f^{-1}_{\rm PBH}).
\end{equation}
Under such assumption, the minihalo can be at most 50 times heavier than the central BH \citep{2019PhRvD.100b3506A}.

\subsubsection{Mini Dark Matter Halo}

The GW emitted by CBC events can also be gravitationally lensed by mini dark matter halos with mass $\sim 10^3-10^6 M_{\odot}$ (hereafter minihalos), with significant cross-section due to wave effect (also diffraction lensing), compared with EM signals. These systems can be viewed as ideal probe to the nature of dark matter, as they are not polluted by sophisticated baryon processes \citep[e.g., ][]{2022PhRvD.106b3018G,2023PhRvD.108j3529T,2024arXiv240805290J,2024PhRvD.109l4020C,2025JCAP...07..025S}. For different dark matter models, the HMF for minihalos given by Equation~\eqref{eq:hmf} may vary. For example, the warm dark matter (WDM) model tends to predict less minihalos at the lower mass end compared with cold dark matter (CDM) model, depending on the particle masses. More importantly, the density profiles of minihalos produced by different dark matter models may be significantly different from each other, especially at the halo centers. Here we make a comparison between the CDM and self-interaction dark matter (SIDM) model as an example, referring the readers to \citet{2024SSRv..220...58V} for extensive discussion on other dark matter models, including WDM and fuzzy dark matter. 

In the CDM framework, minihalos follow the NFW profile (Eq.~\ref{eq:nfw}), which possesses a steep, cuspy inner slope (power-law index $\sim -1$). By contrast, in the SIDM models, frequent particle collisions in the halo center erase the central cusp and produce a nearly constant-density core. In this case, the density profile of a dark matter halo can be divided into two parts, separated by the characteristic radius $r_{1}$, where the average scattering rate per particle times the halo age $t_{\rm age}$ is equal to unity \citep{2016PhRvL.116d1302K}, i.e.,
\begin{equation}
\frac{\langle\sigma_{\rm DM} v\rangle}{m_{\rm DM}}\rho (r_1)t_{\rm age} \sim 1,
\end{equation}
with $\sigma_{\rm DM}$, $m_{\rm DM}$, and $v$ representing the scatter cross section of SIDM particles, dark matter particle mass, and relative velocity of dark matter particles, respectively. For the region with radius $>r_1$, scatter of dark matter particles is inefficient and the density profile is close to the NFW profile, while for that with radius $<r_1$, the density profile may be solved by requiring hydrostatic equilibrium (if ignoring baryons in minihalos), i.e.,
\begin{equation}
\sigma_{\rm DM,0}^2\nabla^2\ln{\rho_{\rm iso}}=-4\pi G \rho_{\rm iso},
\label{eq:sidm}
\end{equation}
with $\sigma_{\rm DM,0}$ representing the one-dimensional velocity dispersion of dark matter particles. Without loss of generality, one can model the SIDM profile as
\begin{equation}
\rho_{\mathrm{SIDM}}(x)=
\begin{cases}
\rho_{\mathrm{iso}}(x), & \mathrm{for~}0<x<r_1, \\
\rho_{\mathrm{NFW}}(x), & \mathrm{for~}x\geq r_1,  
\end{cases}
\end{equation}
where $\rho_{\rm iso}$ is the solution to Equation~\eqref{eq:sidm}.  In the above expression, it is often assumed that the dark matter density and enclosed mass for the inner and outer profiles should match at this special radius $r_1$. Notably, as for the typical cross-section of $\langle\sigma_{\rm DM} v\rangle/{m_{\rm DM}} \sim 1-10^2 \rm cm^2\,km\,s^{-1}\,g^{-1}$, the value of $r_1$  is about $\sim 0.1-2$.

\section{Detection rates of lensed GW events}
\label{sec:rates}

The estimation of the detection rate of the gravitationally lensed GW events is important for the searching strategy preparation in current and future GW detection at different frequencies, from several hundred Hertz to nanohertz. The forecast on the detection of lensed GW events has been intensively discussed, and in general the detection rate is estimated using the \textbf{lensing optical depth} $\tau^{\ell}$, defined as the probability that a GW event at redshift $z_{\rm s}$ is lensed once by line of sight intervening objects/systems, i.e.,
\begin{equation}
\tau^{\ell}=\frac{1}{4\pi}\int_{0}^{z_{\rm s}}dz_{\ell}\int d\boldsymbol{x} {S_{\rm cr}}(\boldsymbol{x},z_{\ell},z_{\rm s})\frac{dn}{d\boldsymbol{x}}\frac{dV_{\rm c}}{dz_{\ell}},
\end{equation}
where $z_{\ell}$ is the redshift and $dn/d\boldsymbol{x}$ denotes the number density distribution of the intervening objects per unit parameters $\boldsymbol{x}$, such as velocity dispersion $\sigma_{\rm v}$ of elliptical galaxies or halo mass $M_{\rm H}$ of the galaxy clusters. The symbol $S_{\rm cr}$ is the lensing cross-section, defined as the area of the caustics in the source plane, i.e., the area within which the lens Equation~\eqref{eq:lens} can have multiple solutions, equivalently multiple images in the geometric-optics regime. In the wave-optics regime, the cross section can be also defined via the detectability of lensed sources \citep[e.g.,][]{2018PhRvD..98j4029D,2022PhRvD.106b3018G}.
With the lensing optical depth, the detection rate $\dot{N}^{\ell}$ can be estimated as
\begin{equation}
%
%\begin{aligned}
%
\dot{N}^{\ell}=\int dz_{\rm s} \int dq\int dm_1 \frac{d^3\dot{N}_{\bullet\bullet}(z_{\rm s})}{dz_{\rm s}dm_1dq}   P(\rho^{\ell}>\rho_0 | m_1,q,z_{\rm s})
\tau^{\ell}(z_{\rm s}).
%
%\end{aligned}
%
\end{equation}
Here $d^3\dot{N}_{\bullet\bullet}(z_{\rm s})/dz_{\rm s} dm_1 dq$ represents the number of GW events per unit time, redshift, primary mass, and mass ratio. The function $P(\rho^{\ell}>\rho_0 | m_1,q,z_{\rm s})$ describes the fraction of lensed events with given source parameters ($m_1,q,z_{\rm s}$) that can be detected considering the S/N enhancement due to the magnification of the signals, also called magnification bias. The symbol $\rho_0$ is the detection threshold for GW signals, which is dependent on both the intrinsic parameters of the GW events and the sensitivity of the GW detectors, for example the noise power spectrum $S_{\rm n}(f)$ of the ground-based GW detectors and the timing-precision $\sigma_t$ of the pulsar timing arrays. Here we should make it clear that the claim of detection of a lensed GW event often involves a de-magnified image, i.e., $\mu^{\ell}<1$. 

The above formalism is also valid for the wave-optics regime, but several key points should be emphasized. As for the diffraction regime, the definition of the cross-section $S_{\rm cr}$ should change for there will be no multiple images. One typical way to define the cross-section within which the diffractive lensing effect to be significant, i.e., the normalized S/N difference $\delta\hat{\rho}$ between lensed and unlensed signals is larger than a certain threshold,
\begin{equation}
\delta\hat{\rho}=\sqrt{\frac{\langle h^{L}-h^{U}_{\rm BF}| h^{L}-h^{U}_{\rm BF}\rangle} {{\langle h^{U}_{\rm BF}}| h^{U}_{\rm BF}\rangle}}>\delta \hat{\rho}_{\rm th},
\end{equation}
where $h^{L}$ is the lensed GW waveform and $h^{U}_{\rm BF}$ is the  best-fit unlensed waveform. In addition, the amplification factor of lensed GW signals in the diffraction regime is frequency dependent and its absolute value is not large, normally $|F(f)|<2$, which highlights the importance of the detection frequency. In this section, we summarize the results in detection rate forecast reported in literature, including the galaxy-CBC, cluster-CBC, dark matter minihalo-CBC, and galaxy-SMBBH lensing systems, where the former and latter two cases are representative of geometric and wave-optics regime respectively.

\subsection{Galaxy/cluster-CBC}

\begin{table*}[h]
\caption{Detection rate of galaxy lensed CBC GW events with different merger rate density evolution, lensing model and GW detectors. 
}
\label{table:rate} 
\centering
\scriptsize
\begin{tabular}{lcccccc} \hline\hline
Source &$\dot{N}^{\ell} [\rm yr^{-1}](R_0 [\rm Gpc^{-3} yr^{-1}])$   &  Channel  & Lensing Model  & Detector& References   \\
        \hline    \hline
\multirow{12}{*}{sBBH} &$0.47-5.05$ (23.9)  &  EMBS \citep{2022arXiv220408732W} & SIE & aLIGO & \citet{2022arXiv220408732W} \\
&$0.84-1.20$ (103)  &  EMBS \citep{2018MNRAS.474.4997C} & SIE+external shear & aLIGO & \citet{2018MNRAS.476.2220L} \\
&$3.5-94.7$ (36)  &  EMBS \citep{2013ApJ...779...72D} & SIS & aLIGO & \citet{2018PhRvD..97b3012N} \\
&$0.58$ (213)  &  EMBS \citep{2017MNRAS.471.4702B} & SIE+external shear & aLIGO & \citet{2018MNRAS.480.3842O} \\
&$2.2-5.0$ (23.9) &  EMBS \citep{2018MNRAS.480.3842O} & EPL & LIGO A+ & \citet{2021ApJ...921..154W} \\
%
%&$38.4 $(103) & EMBS \citep{2018MNRAS.474.4997C} & SIE & LIGO Voyager & \citet{2021PhRvD.103j4055W} \\
%
&$11.2-25.2$ (23.9) & EMBS \citep{2018MNRAS.480.3842O} & EPL & LIGO Voyager & \citet{2021ApJ...921..154W} \\
%        
%&$157.1$ (103) & EMBS \citep{2018MNRAS.474.4997C} & SIE & ET & \citet{2021PhRvD.103j4055W} \\
%
&$2.1-58.8$ (200) &  EMBS \citep{2013ApJ...779...72D} & SIS & ET &  \citet{2015JCAP...12..006D} \\
&$38.6-79.4$ (103) & EMBS \citep{2018MNRAS.474.4997C} & SIE+external shear & ET & \citet{2018MNRAS.476.2220L} \\
&$1120$ (213)  &  EMBS \citep{2017MNRAS.471.4702B} & SIE+external shear & ET & \citet{2018MNRAS.480.3842O} \\
%
%&$184.7 $ (103) & EMBS \citep{2018MNRAS.474.4997C} & SIE & CE & \citet{2021PhRvD.103j4055W} \\
%
&$342.9-437.5$ (200) & EMBS \citep{2013ApJ...779...72D} & SIS & CE+ET & \citet{2022MNRAS.509.3772Y} \\
& $4.5-20.3$ (14) & Dynamical \citep{2021MNRAS.500.1421Z} & SIE+external shear & ET & \citet{2023ApJ...953...36C} \\
& $6.6-29.6$ (14) & Dynamical \citep{2021MNRAS.500.1421Z} & SIE+external shear & CE & \citet{2023ApJ...953...36C} \\    \hline 
\multirow{6}{*}{BNS} &$2.31\times 10^{-4}$ (310) &  \citep{2022MNRAS.509.1557C} & SIE+external shear & LIGO A+ & \citet{2023MNRAS.518.6183M} \\ 
&$1.88\times 10^{-2}$ (310) &  \citep{2022MNRAS.509.1557C} & SIE+external shear & LIGO Voyager & \citet{2023MNRAS.518.6183M} \\ 
&$5.32$ (310) &  \citep{2022MNRAS.509.1557C}  & SIE+external shear & ET & \citet{2023MNRAS.518.6183M} \\ 
&$0.8-9.6$ (100) & \citep{2013ApJ...779...72D} & SIS & ET & \citet{2015JCAP...12..006D} \\ \
&$67.3$ (310) &  \citep{2022MNRAS.509.1557C} & SIE+external shear & CE & \citet{2023MNRAS.518.6183M} \\ 
&$3.88-4.54$ (100) &  \citep{2013ApJ...779...72D} & SIS & CE+ET & \citet{2022MNRAS.509.3772Y} \\
    \hline 
\multirow{2}{*}{BHNS}&$3.40-4.53$ (10)  & \citep{2013ApJ...779...72D} & SIS & ET & \citet{2022MNRAS.509.3772Y} \\
&$7.13-7.23$ (10) & \citep{2013ApJ...779...72D} & SIS & CE+ET & \citet{2022MNRAS.509.3772Y} \\
    \hline\hline
\end{tabular}
\end{table*}

\begin{figure}
\centering
\includegraphics[width=1.0\columnwidth]{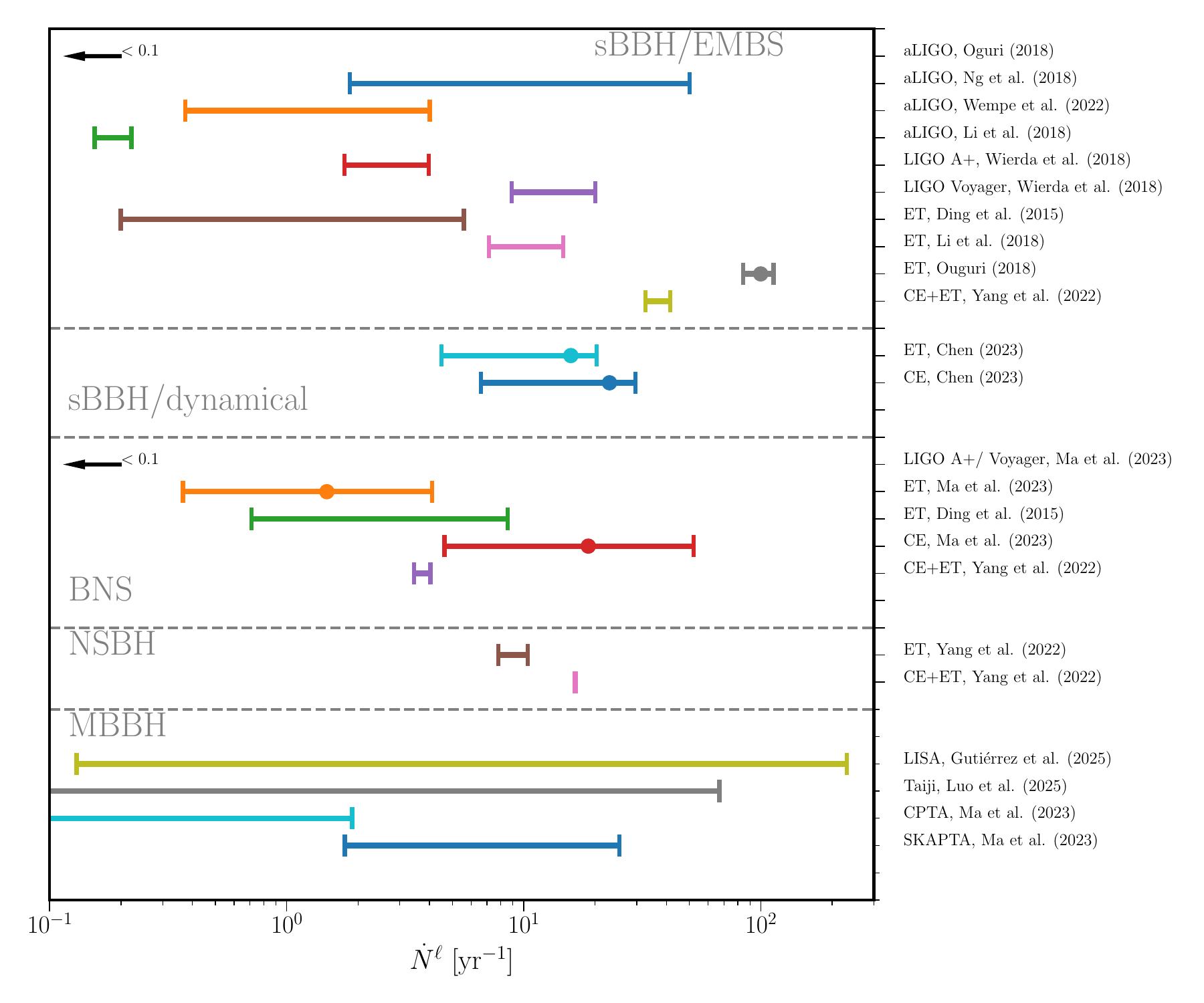}
\caption{
Estimates for the Detection rate of strongly lensed CBC and MBBH systems by different GW detectors.  The results of CBCs are rescaled by the latest constraint on the local merger rate density from GWTC-4.0, i.e., $R^{\rm E}\sim 19_{-5}^{+7} \rm Gpc^{-3} yr^{-1}$ for sBBHs, $89_{-67}^{+159} \rm Gpc^{-3} yr^{-1}$ for BNSs, and $23_{-13}^{+20} \rm Gpc^{-3} yr^{-1}$ for BHNSs \citep{2025arXiv250818083T}.  
}
%Di}
%
\label{fig:merger_rate}
\end{figure}

The frequency of GW emitted by CBCs detected by ground-based GW detectors are in the range from ten Hz to several hundred hertz or even higher. Therefore, the galaxy/cluster lensing is in the geometric-optics regime. As for the galaxy-lensed GW signals emitted by CBCs, the detection rate forecasts for various GW detectors and sources have been intensively discussed in the literature \citep[e.g.,][]{2015JCAP...12..006D, 2018MNRAS.476.2220L, 2018PhRvD..97b3012N, 2018MNRAS.480.3842O, 2021ApJ...921..154W, 2022MNRAS.509.3772Y, 2022arXiv220408732W, 2023ApJ...953...36C, 2023MNRAS.518.6183M}. For a convenient reference for readers, we sort these results in Table~\ref{table:rate}. Since current predictions on the CBC merger rates still have large uncertainties due to that many complicated dynamical and radiative processes involved in the formation of CBCs are not well understood and the contribution fractions from different mechanisms are difficult to accurately estimate, therefore,  it is challenge to estimate the lensed event rate accurately. Recently, the usual approach is to normalize modelled merger‑rate densities to the local rate ($R_0$) inferred from LVK observations while preserving the modelled redshift (or mass) evolution. The latest constraints on the local merger rate densities for sBBHs, BNSs, and BHNSs are $R_0\sim 19^{+7}_{-5} \rm Gpc^{-3}yr^{-1}$, $\sim 89^{+159}_{-67} \rm Gpc^{-3}yr^{-1}$, and $23^{+20}_{-13}\rm Gpc^{-3}yr^{-1}$, respectively, still with quite large uncertainties \citep{2025arXiv250818083T}. As seen from Table~\ref{table:rate}, after rescaling the local merger rate to the same value for each type of CBCs, the predicted values by different authors are different by a factor up to $2-5$ for sBBHs, BNSs, and BHNSs, respectively. Such differences are partly due to the different redshift evolution shapes of the adopted merger rate densities, which now cannot be tightly constrained by the LVK observations due to the lack of GW events detected at high redshifts. For example, \citet{2018MNRAS.476.2220L} found that the choice of different SFR density evolution models (consequently different $R(z)$) can introduce a fractional uncertainty of $50\%-200\%$ to $\dot{N}^{\ell}$ depending on adopted GW detectors. In this case, the higher sensitivity of the adopted GW detector, the larger the fractional uncertainty, simply because more distant sources can be detected by the detector where the differences of $R(z)$ between different models are larger. In addition, the choice of the lensing model also affect the estimate of  $\dot{N}^{\ell}$. For different lensing models, e.g., SIS, SIE, or EPL model, the estimated lensing cross-section $S_{\rm cr}$ may be different by a factor of $\sim 1-5$, which depending on detail settings of the lensing model, e.g., ellipticity for SIE and power law index for EPL. Furthermore, one may also seen from this table, that the detection rate $\dot{N}^{\ell}$ of galaxy lensed BNS mergers may be substantially lower than that of sBBH merger if assuming 2.5G GW detectors (i.e., aLIGO, LIGO A+ and LIGO Voyager), which is due to their significantly lower S/N at the same redshift. However, when it comes to next generation GW detectors, the detection rate of lensed BNS mergers may be larger than that of sBBH mergers due to their relatively larger merger rate density. 

\begin{figure}
\centering
\includegraphics[width=0.8\columnwidth]{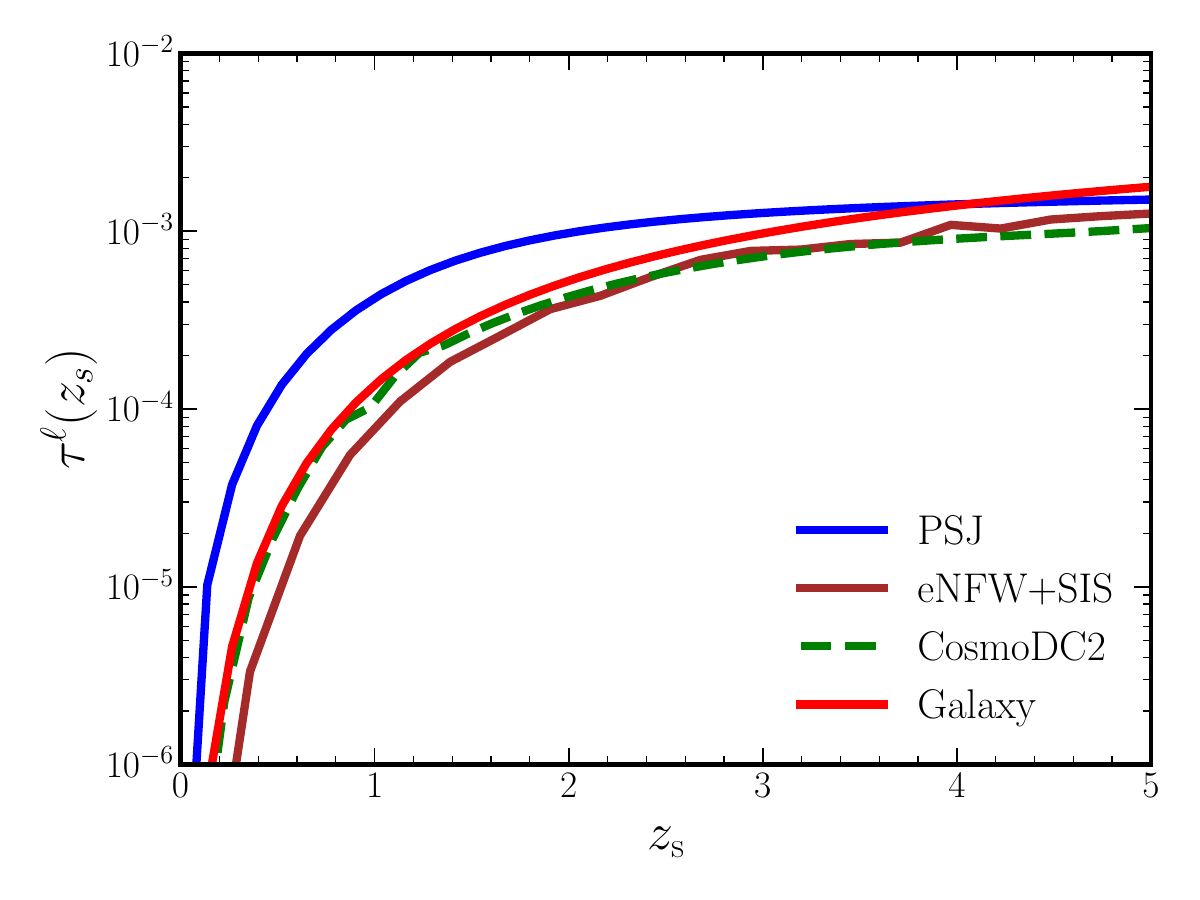}
\caption{The redshift evolution of the lensing optical depth for galaxies (red dashed) and galaxy clusters modeled by various profiles, including the PSJ (blue solid) and eNFW+SIS profile (brown dashed). The green dashed line shows the results from the CosmoDC2 catalog data extrapolation. 
}
\label{fig:tau}
\end{figure}

Although galaxy‑scale lensing of CBCs has been intensively studied, published estimates of detection rates for CBCs lensed by galaxy clusters remain scarce. Though the number density of clusters is much lower than that of galaxies, the cluster-lensing can also be important due to their relatively larger cross-section. \citet{2023MNRAS.520..702S} estimated the total detection rate of lensed GWs for LIGO and Virgo, accounting for both galaxy- and cluster-scale lenses; however, their calculations may be over-simplified because it adopts a universal SIS model to compute lensing cross sections for both galaxies and clusters. A more detailed study of cluster-lensed sBBH mergers for the third-generation GW detectors is given by \citet{2024ApJ...977...64C}, who model galaxy-cluster lenses using mock clusters from the cosmoDC2 catalog or adopt either the pseudo-Jaffe (PSJ) profiles or eccentric NFW halos plus a BCG represented by a SIS profile. Figure~\ref{fig:tau} shows the cosmic evolution of lensing optical depth $\tau^{\ell}$ of both galaxies and galaxy clusters.

The resulted $\dot{N}^{\ell}$ is about $5-84 \rm yr^{-1}$ across all models, and specifically $\dot{N}^{\ell}$ for more realistic galaxy clusters with central main and outer member galaxies in the cosmoDC2 catalog is about $\sim 13 \rm yr^{-1}$, which is almost comparable with that for galaxy-lensed sBBH mergers, i.e., $\sim 14 \rm yr^{-1}$, assuming the same merger rate density evolution and $R_0\sim 19\rm Gpc^{-3}\rm yr^{-1}$. In \citet{2024ApJ...977...64C}, the magnification bias is neglected as almost all sBBH mergers within $z_{\rm s}\in [0,5]$ can be detectable by the third-generation GW detectors with high sensitivity. However, it is different for the cluster-lensing of BNS mergers, of which the S/Ns are relatively low and only those mergers with redshift $z\lesssim 2$ can be detectable, and thus the magnification bias becomes important. We note that estimating the detection rate of lensed BNS mergers by galaxy clusters is especially challenging because magnification bias depends sensitively on the complex mass structure of  clusters, which is hard to be fully considered due to the number and locations of subhalos are difficult to determine. The detection‑rate estimates above should be regarded as optimistic: in realistic searches the identification of lensed CBCs is hampered by false positives and selection effects, which can substantially reduce the effective detection yield. One way to estimate the false alarm probability (FAP) is to calculate the Lens Bayes factor analytically \citep[e.g., ][]{2018arXiv180707062H,2023arXiv230413967G}. For example, \citet{2023arXiv230413967G} showed that $50.6\%$ of the lensed sBBH pairs detected by ET can be identified, while this number rises to $87.3\%$ for the CE+ET network, owing to its superior spatial resolution. Another way is to consider the parameter overlaps of those mock lensed pairs and impostors \citep{2023PhRvD.107f3023C}. The total FAP for the GW lensing detection is defined as the probability of at least one pair within a population of $N$ events can mimic lensing due to astrophysical coincidence, i.e., 
\begin{equation}
{\rm FAP}= 1-(1-{\rm FAP}_{\rm pair})^{N_{\rm pair}},
\end{equation}
where $N_{\rm pair}=N(N-1)/2$ is the total number of event pairs and $\rm FAP_{pair}$ is the percentage of these pairs with parameter overlaps that mimic the lensed pairs. As shown in Table \uppercase\expandafter{\romannumeral3} of \citet{2023PhRvD.107f3023C}, the combined $\rm FAP_{pair}$ is approximately the order of $10^{-5}\sim 10^{-6}$ for the $95\%$ confidence level lensed GW events detection of the LVK network, which is the combination of false alarm due to mass, sky localization, and coalescence phase overlap. Several works indicated that the third generation GW detectors (ET and CE) will constrain source parameters with precision better by factors of a few tens relative to current LVK measurements, thereby shrinking the overlap region in parameter space by a comparable factor \citep[e.g., ][]{2018PhRvD..97f4031Z, 2022NatSR..1217940P}. The substantially increased sensitivity of CE and ET is thus expected to reduce the pairwise FAP, $\rm FAP_{pair}$ to the order of $10^{-8}-10^{-10}$ \cite{2023ApJ...953...36C}. Then the total FAP for the lensed GW events detected by the third generation detectors is on the order of $10^{-2}-10^{0}$ per year, which is significantly smaller than the predicted number of the detection year rate shown in Table~\ref{table:rate}. For similar reasons, \citet{2025arXiv251023238B} claimed that the improved parameter precision will indeed compensate for the growth in false alarms in the O5 run of LVK, making the first $3\sigma$ detection come true. In addition, the micro-lensing of stars in galaxies or galaxy clusters may also help to more easy identification of lensed GW signals, since it may change the waveform of lensed GW signals differently at different frequencies \citep[e.g,][]{2024SCPMA..6769511S, 2023arXiv230106117S}. Therefore, we are optimistic on the detection of lensed GW events in the upcoming era of the LIGO+, LIGO Voyager and third generation ground-based detectors.

\subsection{Dark matter minihalo/PBH-sBBH}

\begin{figure}
\centering
\includegraphics[width=0.8\columnwidth]{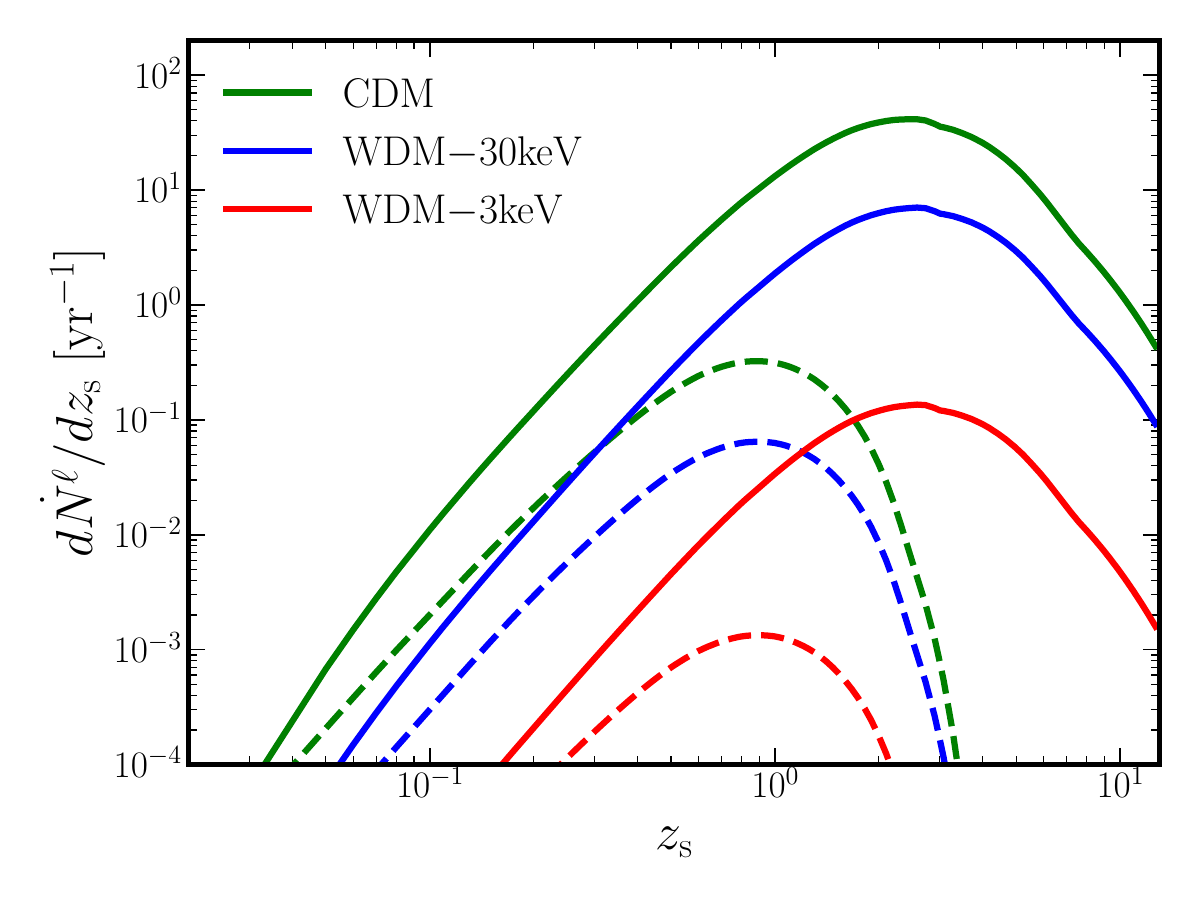}
\caption{The differential detection rate of sBBH mergers diffractively lensed by minihalos with LIGO A+ (dashed) and CE (solid), assuming CDM (green), WDM-30keV (blue) and WDM-3keV (red) models. 
}
\label{fig:dndz}
\end{figure}

We summarize the estimates for the detection rates of sBBH mergers lensed by dark matter minihalo and PBH in this section. These lens systems are in the wave-optics regime. Figure~\ref{fig:dndz} shows the differential detection rate of sBBH mergers lensed by different dark matter models. As for dark matter minihalo lenses, if adopting the CDM model (NFW density profile for lens) and an S/N difference threshold of $\delta\hat{\rho}_{\rm th}=1$, then the detection rate is about $\dot{N}^{\ell}\sim 6.44\times 10^{-7}$, $4.21\times 10^{-3}$, $0.288$, $1.02$, $9.76$, and $189~\rm yr^{-1}$ for the high-frequency ground-based GW detectors LIGO A+, ET, CE, Gravitational-wave Lunar Observatory for Cosmology (GLOC), and future middle-frequency space-based DECIGO and BBO, respectively \citep[see][]{2022PhRvD.106b3018G}. 
As for the SIDM model, the detection rate is strongly dependent on the size of the core of the lenses. For example, if the core size is rather small, e.g., $r_1=0.1$, the detection rate $\dot{N}^{\ell}$ is more or less similar with that of CDM model. However, if we impose a large core, for instance $r_1=1.0$, the value of $\dot{N}^{\ell}$ will be substantially suppressed, i.e., to $\sim 2.45\times 10^{-9}$, $8.93\times 10^{-6}$, $4.69\times 10^{-4}$, $2.37$, and $88.2~\rm yr^{-1}$, correspondingly. As for the WDM model, the detection rate is dependent on the particle mass. For example, assuming the particle mass of $30\rm keV$ and $3\rm keV$, the detection rates $\dot{N}^{\ell}$ are about $\sim 0.09$ and $0.002$ per year for CE. Apparently, the middle-frequency GW detectors, such as DECIGO and BBO, have sufficiently high sensitivity to detect lensed GW events via diffractive lensing by minihalos and help to reveal the nature of dark matter, while it is challenging for the current high-frequency ground-based GW detectors to detect such lensing effect. As for fuzzy dark matter model with axion mass of $m_{\rm a}\sim 10^{-17}~\rm eV$, \citet{2023JCAP...07..007F} estimated that the detection rate of lensed events by ET is about $5\times10^{-4}-4\times10^{-3}$ and $0.01-0.07$ per year. If the axion mass is substantially low, i.e., $m_{\rm a}\lesssim 10^{-18}~\rm eV$, then the detection of such lensing effect will be extremely difficult. In conclusion, the dark matter nature, either cold, warm, or self-interacting, may be constrained via the detection of the diffractively lensed GW events by future GW detectors.

The detection rate of sBBH mergers lensed by PBHs are rarely discussed in the literature, especially those by bare PBHs \citep[][]{2024PhRvL.133j1002C,2025PhRvD.112d3038L}. However, we comment that such cases may act similarly as stars in galaxies and only change the waveform slightly. As for the dressed PBH, the detection rate of lensed sBBH mergers has been used to constrain the PBH mass $m_{\rm PBH}$ and abundance $f_{\rm PBH}$. In \cite{2023PhRvD.108b3507U}, they find that $\dot{N}^{\ell}$ by LVK detection is strongly dependent on $f_{\rm PBH}$ and $m_{\rm PBH}$. For example, if $m_{\rm PBH}>100M_{\odot}$ and $f_{\rm PBH}=1$, the detection rate is about $\dot{N}^{\ell}\sim 1.75 \rm yr^{-1}$. On the other hand, the value of $\dot{N}^{\ell}$ is about $0.5 \rm yr^{-1}$ for $f_{\rm PBH}=0.2$ and $m_{\rm PBH}\sim 20 M_{\odot}$. This result is easy to understand, for more massive PBHs can have larger cross-section and the larger abundances the larger lensing optical depth. The above results indicate that the existing null-finding of sBBH mergers lensed by PBHs disfavors  $f_{\rm PBH}>0.7$ at $1\sigma$ in the mass range $m_{\rm PBH}>50 M_{\odot}$. This constraint power can be further improved by next-generation GW detectors like ET. 

\subsection{MBBH-Galaxy}

Inspiraling and merging SMBBHs/MBBHs can also be lensed by intervening galaxies and galaxy clusters, with optical depth determined by the distributions of these lenses and the lensing cross-section. The detection of such events along with their EM counterparts (also lensed host galaxies) can also applied for probing cosmological parameters and test general relativity, etc.

The nanohertz GW signals from SMBBH inspirals can be gravitational lensed by foreground galaxies in the wave-optics regime, which may be observed by future PTAs. By taking into account the diffraction effect, \citet{2023MNRAS.524.2954M} investigated the GW signals from the lensed SMBBHs and estimated the detectable number of such signals by future PTAs based on the population synthesis model of SMBBHs described in Section~\ref{sec:MBBH}. They found that the lensed GW signals are only amplified by a factor slightly larger than $1$ with slightly shifted phases and the expected detection number is on the order of hundreds with $30~\rm yrs$ of observation by CPTA and SKA-PTA. Though the expected number appears optimistic for detection, the identification of such lensed GW signals is extremely challenging if they are from circular SMBBHs as monochromatic sources, because the lensed waveforms cannot be distinguished from un-lensed signals. One possible way to identify such events is to monitor their associated lensed host galaxies (non-active SMBBH case) or lensed AGNs (active SMBBH case). Here we also comment that if the orbit of a lensed SMBBH is elliptical, the lensing effect can be much more significant. This is because the GW waveform in this case is the superposition of different harmonics with different frequencies. For different harmonics, the amplification factor is different, which leads to frequency-dependent changes of the waveform that may not be matched by un-lensed signals. By extracting such frequency-dependent amplification feature, one may identify the lensing nature of the source.

Note here that \citet{2023ApJ...955...25K} treated such lensed systems in the geometric-optics regime and predicted the number potentially detectable. Their estimates may be optimistic, however, because magnification factors in the geometric-optics regime are generally larger than those obtained in the wave-optics regime. 

On the one hand, for MBBH merger systems emitting millihertz GWs, the lensing by galaxies is in the geometric-optics regime \citep[e.g.,][]{2010PhRvL.105y1101S}. By taking galaxies as intervening lens, \citet{2025arXiv251002061G} and \citet{Luo2026} predicted that LISA and Taiji can detect about $\sim 0.04-231$ strongly lensed MBBH mergers in a four-year observation period (see also Figure~\ref{fig:merger_rate}), depending on different source populations, including light- and heavy-seed scenarios. On the other hand, MBBH mergers may also be lensed by low-mass DM halos or subhalos, which belongs to the wave-optics regime for their smaller sizes \citep{2023PhRvD.107d3029C}.  By considering the diffraction waveform including merger, ring-down and higher terms, \citet{2023PhRvD.108l3543C} systematically estimate the detection rate of MBBH mergers with diffractive lensing signatures  for LISA, which is $\sim 0.01-7.96$ for $1\sigma$ detection threshold. Notably, these rates are strongly subject to the DM HMF, such as the theory-based Press-Schechter halo mass function and an observation-based model derived from Sloan Digital Sky Survey they adopted and similarly also MBBH populations.

\subsection{Other systems}

Here in this section, we briefly summarize the detection rate estimation of some other lensed GW systems, including the lensed EMRI/IMRI systems detected by space-based mHz detector such as LISA, Taiji, and Tianqin, lensed inspiraling double compact objects and GW from Tidal disruption event detected by middle frequency detectors such as DECIGO. 

\begin{itemize}
\item[(1)] \textbf{Lensed EMRIs/IMRIs}: One of the major scientific goal of space-based GW detectors like LISA, Taiji, and Tianqin is to detect the inspiral of a $\sim 1-100 M_{\odot}$ compact object into a $10^4-10^7 M_{\odot}$ MBH, which is denoted as EMRIs if mass ratio $q\sim 10^{-5}-10^{-7}$ and IMRIs if mass ratio $q\sim 10^{-2}-10^{-4}$. These sources are natural excellent laboratories for testing strong-field gravity. Considering lenses with masses $\gtrsim 10^7 M_\odot$, \citet{2024PhRvD.109f3505T} found that the lensing of EMRIs belongs to the geometric-optics regime and the detection rate of EMRIs by LISA is about $0-40$ over a mission time of four-years, depending on the EMRI population.  
\item[(2)] \textbf{Lensed GW from TDE}: Tidal disruption event is a phenomenon caused by a star moving too close to a MBH thus being torn apart and accreted by the MBH. Not only the EM signals of a small fraction of TDEs can be lensed \citep[e.g., ][]{2024ApJ...962....3C,2024A&A...690A.384S}, but also the GW emitted from it can be lensed and possibly detected by deci-Hz detectors such as DECIGO. \citet{2023MNRAS.523.3863T} considered the GW signals from TDEs lensed by galaxies with masses of $10^8-10^{12} M_\odot$ in the geometric-optics regime. They found that with an observation period of $10$\,yrs, DECIGO can detect $\sim 10$ strongly lensed nuclear TDEs (main sequence star disrupted by MBHs in galactic centers) and $\sim 130$ off-nuclei TDEs in globular clusters (white dwarfs disrupted by intermediate MBHs in globular clusters). These multi-messenger TDE events can serve as important probes to constrain the properties of TDEs. 
\item[(3)] \textbf{Lensed Inspiraling binary compact objects}: Not only the GW produced by CBCs can be lensed, but also the GW produced in their inspiral stage can be lensed and detected by deci-Hz GW detectors like DECIGO \citep[e.g.,][]{2021PhRvD.103d4005H,2021ApJ...908..196P,2023MNRAS.523.3863T}. For example, \citet{2023MNRAS.523.3863T} investigated the inspiraling double compact objects lensed by intervening galaxies in the geometric-optics regime, and found that the expected numbers are $\sim 0-0.29/0.02-0.39/8.46-71.28$ for BNS, BHNS and sBBH cases within an observation period of $4$\,yrs. 
\end{itemize}

\section{Cosmological Application }
\label{sec:application}

\begin{figure}
\centering
\includegraphics[width=1.0\columnwidth]{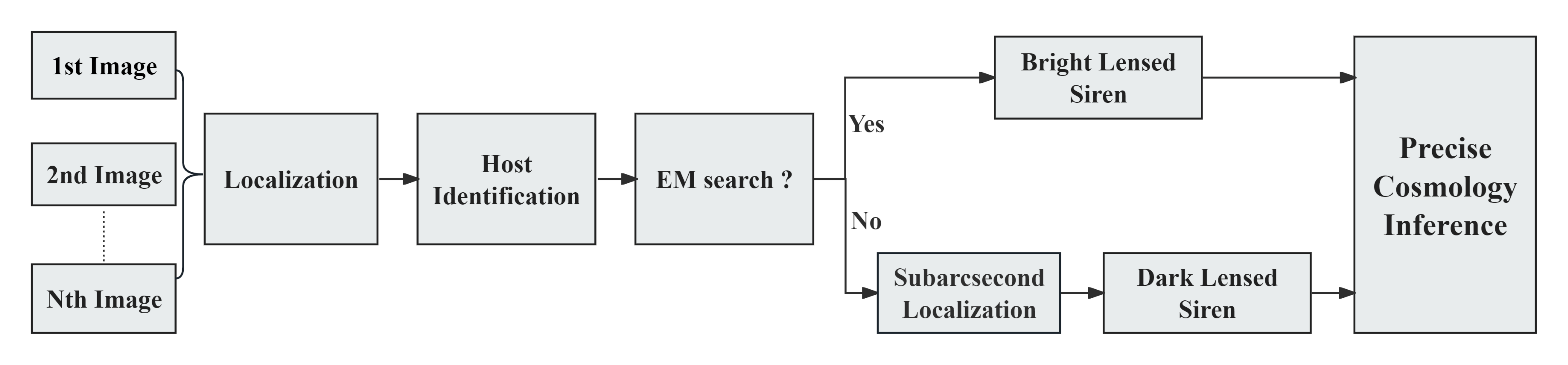}
\caption{Steps for precise cosmological inference via the strongly lensed gravitational wave signals. 
}
\label{fig:step}
\end{figure}

The gravitational lensing of GWs holds a wealth of pivotal applications in astrophysics, including the following:
(1) Probing the nature of dark matter through distinct GW waveform features and detection rates produced by lensing from intervening dark matter halos with varying intrinsic properties \citep[e.g., ][]{2018ApJ...867...69L,2020MNRAS.495.2002L,2021MNRAS.502L..16C,2022PhRvD.106b3018G,2022ApJ...926L..28B,2022A&A...659L...5C, 2023PhRvD.108j3529T,2024arXiv240805290J, 2024PhRvD.109l4020C,2024ApJ...975...48B};  
(2) Constraining the mass and cosmic abundance of PBHs via lensing signatures in the waveforms of point-like GW sources and the distortion of the stochastic gravitational wave background \citep{PhysRevLett.122.041103, 2020PhRvD.101l3512D, 2021PhRvD.104h3515W, 2023PhRvD.108b3507U, 2025arXiv251013477S};
(3) Constraining the astrophysical origin and formation channels of sBBH mergers by cross-correlating lensed GW events with their host galaxies  \citep[e.g.,][]{2022ApJ...940...17C,2022MNRAS.517.4656H,2023ApJ...953...36C}; 
(4) Placing tight constraint on the GW propagation speed by comparing the arrival times and time-delays between lensed GWs and their EM counterparts \citep{2017PhRvL.118i1102F,2017PhRvL.118i1101C};
(5) Setting limits on the graviton mass and probing the nature of gravity via the time-delays and waveform-morphology modifications induced by gravitational lensing \citep[e.g.,][]{2019ApJ...880...50Y,2019ApJ...875..139L, 2020PhRvD.101j3509M, 2021PhRvD.104l4060C,Wright:2024mco}. 

%from probing the nature of dark matter \citep[e.g., ][]{2018ApJ...867...69L,2020MNRAS.495.2002L,2021MNRAS.502L..16C,2022PhRvD.106b3018G,2022A&A...659L...5C, 2023PhRvD.108j3529T, 2024arXiv240805290J, 2024PhRvD.109l4020C,2025arXiv251013477S}, constraining the mass and abundance of PBHs \citep[e.g.,][]{PhysRevLett.122.041103,2020PhRvD.101l3512D, 2021PhRvD.104h3515W, 2023PhRvD.108b3507U} and the origin of sBBH mergers \citep[e.g.,][]{2022ApJ...940...17C,2022MNRAS.517.4656H,2023ApJ...953...36C}, limiting the GW propagation speed \citep[e.g.,][]{2017PhRvL.118i1102F,2017PhRvL.118i1101C}, to limiting the mass of graviton or nature of gravity \citep[e.g.,][]{2019ApJ...880...50Y, 2020PhRvD.101j3509M, 2021PhRvD.104l4060C}. 
Furthermore, the gravitational lensing of GWs can also be applied as an independent probe to constrain cosmological parameters \citep[e.g., ][]{Liao=2017,2019ApJ...873...37L, 2020PhRvD.101f4011H, 2021MNRAS.507..761H, 2020MNRAS.498.3395H, 2023PhRvL.130z1401J,2025arXiv251215168M,zhiwei_na}, especially it may be helpful in resolving the Hubble tension. In this section, we outline two methods to infer the cosmological parameters, namely ``Bright Lensed Siren'' \citep{2017NatCo...8.1148L} and ``Dark Lensed Siren'' \citep{zhiwei_na}. In general, both methods take the advantage of the extremely precise measurement on the time-delay $\Delta t_{i,j}$ among lensed images of GW signals,
\begin{equation}
\Delta t_{i,j}=\frac{D_{\rm l}D_{\rm s}}{D_{\rm ls}}\frac{(1+z_{\rm l})}{c}\Delta\phi_{i,j},
\end{equation}
where $\Delta\phi_{i,j}$ represents the lensing potential difference between the $i$-th and $j$-th images, which can be reconstructed by EM observation, i.e., lensed EM counterparts (bright lensed sirens) or lensed host galaxies (dark lensed sirens). Then the cosmological parameters can be constrained by the combination of the angular diameter distances, i.e., time-delay distance $D_{\tau}={D_{\rm l}D_{\rm s}}/{D_{\rm ls}}$.

\subsection{Bright Lensed Siren}

The ``Bright Lensed Siren'' was first proposed by analogizing to the time-delay cosmography method of strongly-lensed Quasars \citep[e.g.,][]{2020MNRAS.498.1420W}. In this method, one may reconstruct the  map of $\Delta\phi$ from observations of lensed host galaxies and then determine the specific  $\Delta\phi_{i,j}$ for the $i$-th and $j$-th images of the lensed GW signals by the positions of lensed EM counterparts, i.e., kilonovae and afterglows, in the host galaxies. The advantage of this method compared with the traditional time-delay cosmography method lies within the following two aspects. First, the time resolution of GW detectors can be extremely precise, normally at the scale of several ms, which is negligible comparing with time-delays between different images (typically days to months). However, the relative uncertainty of time-delay measurement for lensed Quasars is normally on the order of $\sim 3-5\%$. Second, comparing with the case of lensed Quasars, the potential map $\Delta\phi$ can be reconstructed more precisely as there is no contamination from AGN light. By performing Monte Carlo Simulations, \citet{2017NatCo...8.1148L} found that using about $\sim 10$ such lensed GW+EM systems can provide a $\sim 0.68\%$ uncertainty of the Hubble constant $H_0$ with next-generation GW detectors.

Note that in the above ``Bright Lensed Siren'' method requires observed EM counterparts of the lensed events to locate the  GW source in its host galaxies \citep[e.g.,][]{2025arXiv251210344L} and thereby measure the specific potential difference $\Delta\phi_{i,j}$ between images. Consequently, this approach applies primarily to BNS and BHNS systems, since sBBH mergers typically lack observable EM counterparts, unless they occur in special environments (e.g., AGN disks). The method's effectiveness therefore depends strongly on the joint GW-EM detection rate. \citet{2023MNRAS.518.6183M} and the follow-up studies by \citet{zhiwei_bnslens} outlined searching strategies for lensed multi-messenger BNS signals and emphasized the importance of identifying lensed host galaxies. By this strategy, the joint GW-EM detection rate for lensed systems can be estimated as
\begin{equation}
\dot{N}^{\ell}_{\rm GW+EM}=\dot{N}^{\ell}_{\rm GW}\times f_{\rm EM}\times f_{\rm Host}  P_{\rm Host}.
\label{eq:flen}
\end{equation}
Here, $\dot{N}^{\ell}_{\rm GW}$ denotes the rate of lensed GW events, which is estimated to be on the order of one to several tens per year for the third-generation GW detectors such as CE and ET (see Table~\ref{table:rate}); $f_{\rm EM}$ represents the fraction of lensed GW events that have detectable lensed EM counterparts, which is set largely by the limiting flux (or magnitude) of a telescope; for example, adopting the infrared band F158 of the Roman Space Telescope (RST), yields $f_{\rm EM}\sim 0.1$ for kilonovae and $\sim 0.01$ for afterglows; $f_{\rm Host}$ denotes the fraction of host galaxies that can be identified as lensed systems in a given survey over a certain sky area, which typically ranges from $\sim 20\%$ to $50\%$, depending on the telescope’s limiting magnitude and the criteria adopted for lens identification; $P_{\rm Host}$ represents the probability of successfully matching the associated lensed host galaxies, depending directly on the localization uncertainty, and $P_{\rm Host} \sim 0.1$ for a localization precision of $1$\,deg$^2$ and $\sim 1$ for $0.1$\,deg$^2$ if assuming a random matching strategy. By detailed modeling of the EM population of BNS mergers, utilizing both the results from observation and numerical simulations to estimate those terms above, \citet{zhiwei_bnslens} found that the joint detection of lensed GW with kilonova is promising for the Roman Space Telescope (RST) and the detection rate $\dot{N}^{\ell}_{\rm GW+K}$ is about $\sim  0.45_{-0.34}^{+0.80}$, $0.55_{-0.41}^{+0.98}$, and $0.06_{-0.05}^{+0.11}$\,yr$^{-1} $ with the RST F106, F158, F213 bands, respectively. As for other telescopes such as CSST and Euclid, the expectation detection rate may be $\sim 10$ times smaller, limited by their small limiting magnitude \citep{2023MNRAS.518.6183M}. The joint detection of lensed afterglow signals in the optical and radio bands is however even more difficult, even with the RST and Square Kilometer Array (SKA). Therefore, achieving a $1\%$ precision measurement of the Hubble constant may require waiting $10-20$\,yrs after third generation GW detectors come online.

\subsection{Dark Lensed Sirens}

Unlike the ``Bright Lensed Siren'', the ``Dark Lensed Siren'' does not rely on EM counterparts and therefore applies to sBBH mergers as well as distant BNS/BHNS merger events whose faint EM emission is undetectable. This increases its efficiency for two reasons. First, it includes lensed sBBH (which typically lack EM counterparts) and BNS/BHNS mergers whose EM signals are too faint to observe, thus substantially enlarge the usable sample. Second, while the local merger rate density of BNS (and BHNS) merger rate density remains uncertain and could be low, the local sBBH merger rate density is relatively well constrained at $\simeq 19_{-5}^{+7} \rm Gpc^{-3} yr^{-1}$ thus the number of detectable BNS mergers could be small. However, the local sBBH merger rate density is relatively well constrained at $\simeq 19_{-5}^{+7}$\,Gpc$^{-3}$\,yr$^{-1}$; combined with their larger masses of sBBHs, this makes nearly all lensed sBBH mergers detectable by third-generation GW detectors, so sBBHs are likely dominate the applicable sources. 

The crucial element of this ``Dark Lensed Siren'' method is identifying and reconstructing the lensed host galaxy, then achieving the sub-arcsecond re-localization of the GW source using GW-measured the time-delay and magnification ratios. There are three necessary key ingredients for the application of this method \citep[see details in][]{zhiwei_na}:
\begin{itemize}
\item {\bf precise localization of the lensed events by GW signals:} The localization of lensed GW events can be achieved by an effective network formed when multiple images of the same event are detected at different times and different Earth locations in the space. This effective network increases the number of independent ``detectors'' and introduce varying antenna pattern functions through Earth motion and orbital rotation, as well as extra baselines between detectors that record different image pairs; these effects together can lead to an improvement of localization by roughly a factor of $30$.
\item {\bf identification of the lensed host galaxy in the localization area:} The typical localization area for a lensed event using the effective network is $\sim 0.01$\,deg$^2$. Within such an area, large scale sky surveys (such as Euclid, Chinese Space Survey Telescope (CSST) \citep{2026RAA....26b4004W}, and RST) would typically found on the order of $0.1$ candidate lensed galaxies. Considering $f_{\rm host} \sim 20\%-50\%$, if a lensed galaxy lies in the GW localization region, it is likely to be the true host. Nonetheless, an FAP of $\sim 1-20\%$ can persist because of Poissonian chance matches, which can be reduced by adding discriminating information of tightening localization constraints.
\item {\bf sub-arcsecond relocalization of the lensed GW event:} With three or more images, the lensed GW source can be relocalized to sub-arcsecond precision by reconstructing the lensed host galaxy using the image's time-delays and magnification ratios. This reconstruction breaks the positional degeneracy that arises for non-spherical lensing models; the central image plays a key role in resolving the degeneracy. The central image should be detectable by third-generation GW detectors due to thier high sensitivity. In contrast to EM observations--where the central image is often heavily demagnified ($\mu\sim0.01$) and contaminated by  the lens galaxy's central light--the GW central image is mildly de-magnified (strain reduced by  $\sqrt{\mu}\sim0.1$) and is free from such contamination (see Section~\ref{sec:galaxy}). 
\end{itemize}
Based on the above recipes, \citet{zhiwei_na} estimated the constraining power of the ``Dark Lensed Siren'' on the Hubble constant, and they found that $H_0$ can be constrained with $\lesssim 1\%$ precision with a period of  about two years detection of the dark lensed events by third-generation GW detectors.

To conclude our discussion of cosmological application of lensed GW signals, we briefly review the current Hubble-constant tension. Local distance-ladder measurements (Cepheids + Type Ia supernovae) give  $H_0=73.01\pm 0.99$\,km\,s$^{-1}$\,Mpc$^{-1}$ \citep[e.g.,][]{2022ApJ...938...36R}, while CMB analysis that combine temperature, polarization, and lensing reconstruction yield $H_0=67.43\pm 0.49$\,km\,s$^{-1}$\,Mpc$^{-1}$ \citep{2020A&A...641A...5P}, a discrepancy at at least $5\sigma$ level, which is currently hard to resolve. Both the ``Bright Lensed Siren'' and ``Dark Lensed Siren'' approaches will provide cosmological constraints that are independent of the aforementioned probes, and thus make a critical contribution to resolving the Hubble-constant tension. Because lensed GW events probe higher redshifts than most local distance indicators, they can also deliver independent information on other cosmological parameters. These properties make lensed-GW methods a promising and complementary tool for next-generation cosmological studies.

\section{Searching in GWTC-data}
\label{sec:search}

Lensed GW events are one type of the main targets for GW detection, which have been searched for in the LVK data from the beginning of the ground-based GW observations. One motivation is that lensing magnification can make intrinsically lower-mass, higher-redshift GW events appear as heavier systems, leading to the proposal that some high-mass BBHs observed by the LVK could be lensed events \citep{2021PhRvD.104j3529D}. Although this possibility illustrates the importance of dedicated lensing analyses, recent studies find that lensing magnification cannot consistently explain the observed high-mass BBH population \citep{2026arXiv260414247H}. More generally, there is still no compelling evidence for lensing signatures, i.e., strong lensing with multiple images or diffractive microlensing with frequency-dependent amplification, in the GW signals observed by the LVK network \citep[e.g.,][]{2021ApJ...923...14A, 2024ApJ...970..191A,2025arXiv251216347T}. Nevertheless, a number of methods have been proposed and applied to search for such lensed events in real data. Here we summarize the main methods currently applied in searching for lensed GW events.

\subsection{Preliminary Identification of Lensed Pair Candidates}

In the geometric-optics regime, the strong lensing of a GW signal produces multiple images, and the extracted physical and geometric properties from different images, such as the localization area, should be more or less similar. Therefore, a natural, simple, and economic method is to conduct posterior-overlap analysis for the preliminary identification of lensed pair candidates. Two statistics commonly used to rank and select initial candidates for lensed multiple images are the overlap Bayes factor $\mathcal{B}^{\rm overlap}$ and the time-delay coherence ratio $\mathcal{R}^{\rm gal}$. The overlap Bayes factor is defined as
\begin{equation}
\mathcal{B}^{\rm overlap}=\int d\Theta \frac{p(\Theta | d_1)p(\Theta | d_2)}{p(\Theta)},
\end{equation}
where $P(\Theta | d_1)$ and $P(\Theta | d_2)$ are the posteriors for the intrinsic and extrinsic parameters ($\Theta$) of  two GW events (or lensed GW images), with $d_1$ and $d_2$ representing the observations of events 1 and 2, respectively. The higher $\mathcal{B}^{\rm overlap}$, the two events are more likely to have similar parameters and therefore be the lensed multiple-images of a single event. The time-delay coherence ratio is defined as
\begin{equation}
\mathcal{R}^{\rm gal}=\frac{p(\Delta t | \mathcal{H_{\rm SL}} )}{p(\Delta t | \mathcal{H_{\rm U}} )},
\end{equation}
where $p(\Delta t | \mathcal{H_{\rm SL}} )$ and $p(\Delta t | \mathcal{H_{\rm U}} )$ are the prior probabilities of time-delay $\Delta t$ under the strongly lensed and unlensed hypotheses, respectively. Note that $\mathcal{R}^{\rm gal}$ is dependent on the lensing model and therefore subject to some astrophysical uncertainties. \citet{2021ApJ...923...14A} selected most promising events pairs by the above quantities, i.e., $\mathcal{B}^{\rm overlap}>50$ (and $\mathcal{R}^{\rm gal}>0.01$ ). In addition, with the rapid development of the machine-learning (ML) technique, it is possible to use ML to classify candidate pairs of lensed events, which is independent of the posterior-overlap analysis \citep{PhysRevD.104.124057}. By ML approach, one did not need to generate the above Bayesian posterior which may take hours to days, but directly construct the time-frequency and localization sky maps in the order of seconds and thus may be used to conduct preliminary identification efficiently. Note here that we also comment that recently \citet{2025ApJ...980..258B} developed a modified version of the posterior overlap method and find it is possible to identify $\sim 65\%$ lensed events with LVK efficiently.

Note that unlensed events can masquerade as lensed ones, i.e., coincident overlaps in inferred parameters, combined with estimation uncertainties can produce comparably high values of the overlap Bayes factor $\mathcal{B}^{\rm overlap}$. For example, \citet{2023PhRvD.107f3023C} found that the FAPs from chance overlaps in chirp mass, sky position, and coalescence phase for current LVK detected events are on the order of $1-10\%$ per pair, implying that the accidental matches may outnumber true lensed pairs. This situation should improve with third-generation GW detectors, which are expected to yield substantially better parameter estimation and thus reduce coincidental overlaps.
With the preliminary identification of lensed pair candidates, despite a nonzero FAP, one can further carry out detailed joint parameter-estimation (PE) analysis on the lensed-pair candidates. 

\subsection{Joint Parameter Estimation Analysis}

Unlike the preliminary identification, the Joint PE analysis operates directly simultaneously on the detector strain data instead of on independently derived posterior samples for each component of the candidate pair. Although computationally intensive, the Joint PE analysis explores the full parameter space and therefore yields substantially stronger tests on the lensing hypothesis than methods that marginalize over or treat parameters separately. There are three estimators in such analysis typically used to quantify the evidence for the strong lensing hypothesis as follows \citep[e.g.,][]{2021ApJ...923...14A,2023PhRvD.107l3015L,2024ApJ...970..191A}. (1) {\bf Coherence ratio $\mathcal{C_{\rm U}^{\rm L}}$:} defined as the ratio of lensed and unlensed evidences, neglecting selection effects and using default priors, which quantifies the confidence of parameters of the GWs aligning with the expectations of lensed GW events, but can not account for the coincident parameter overlap. (2) {\bf Population-weighted coherence ratio $\left.\mathcal{C_{\rm U}^{\rm L}}\right|_{\rm pop}$:} taking into account the likelihood that GW parameters overlap by chance. However, the value of $\left.\mathcal{C_{\rm U}^{\rm L}}\right|_{\rm pop}$ is largely subject to the choice of both CBC population and lens models. (3) {\bf Bayes factor $\mathcal{B}_{\rm U}^{\rm L}$:} quantifying the evidence for the strong lensing hypothesis given a detector network and the population model, which explicitly takes into account the selection effects (for example, certain CBC mass ranges and sky orientations are preferentially detected because of the GW detectors' PSDs). Among these three estimators, the most decisive  is the Bayes factor $\mathcal{B}_{\rm U}^{\rm L}$, which is however computationally expensive. A common procedure is to first select pair candidates with high $\mathcal{C_{\rm U}^{\rm L}}$ and high $\mathcal{C_{\rm U}^{\rm L}}\mid_{\rm pop}$ (for example, $\log \mathcal{C_{\rm U}^{\rm L}}>4$), and then compute $\mathcal{B}_{\rm U}^{\rm L}$ only for these candidates. 

The above joint PE procedure has been implemented into pipelines, such as \texttt{GOLUM} \citep{2023MNRAS.526.3832J} and \texttt{HANABI} \citep{2023PhRvD.107l3015L}. Both the pipelines adopt the nested sampling algorithm \texttt{Dynesty} \citep{2020MNRAS.493.3132S} and implement the joint PE with the help of \texttt{Bilby} \citep{2019ApJS..241...27A,2020MNRAS.499.3295R}. In a recent paper by \citet{2024ApJ...970..191A}, \texttt{HANABI} was adopted to estimate $\mathcal{B}_{\rm U}^{\rm L}$ for $17$ pairs, all of which yielded $\mathcal{B}_{\rm U}^{\rm L}<1$. Therefore, it still remains no evidence for lensing signatures in current GWTC data. Notably, \citet{2023MNRAS.526.3832J} performed a systematic follow-up analyses of the O3 LVK lensing searches and identified two strong-lensing candidate pairs, i.e., GW191103-GW191105 and GW191230\_180458-GW200104\_180425, but were unable to obtain high statistical significance for lensing hypothesis. 

\subsection{Search for Sub-threshold Lensed Images}

In the above methods, the GW event pairs should have S/N above the detection threshold, such that they can be marked down and put into the GWTC catalog. However, due to the de-magnification of the gravitational lensing effect, some of the images may not be claimed as detection for its low S/N. Therefore, it is also important to search for GW pairs with sub-threshold lensed images, which may be buried within noise. It is practical to assume that lensed GWs, no matter super-threshold or sub-threshold, originating from the same source, share the same intrinsic masses and spins. Consequently, one can construct a reduced template bank, comprising templates whose masses, spins, and other relevant properties match those of a target super-threshold detection. Given these templates, it is possible to search for potential sub-threshold lensed images matching each target super-threshold GW events existed in the GWTC catalog. Then, we may go through the above procedure to verify the possibility of the candidate pairs to be lensed GW signals. \citet{2024ApJ...970..191A} found that the top 5 ranking candidates from sub-threshold searches all have coherence ratio $\mathcal{C_{\rm U}^{\rm L}}<1$ with all the super-threshold events. Therefore, no compelling evidence for sub-threshold lensed images have been identified. 

\subsection{Search for Microlensing Effect}

The microlensing of GW signals alters the waveform not only through simple phase shifts and amplitude magnification but also via frequency-dependent diffractive effects. However, these modifications depend sensitively on the lens model. To keep searches computationally tractable, the microlensing is therefore commonly approximated as arising from isolated point masses. In this case, the microlensed waveform is
\begin{equation}
h^{\rm M}(f;\theta,M_{\rm L}^{z},y)=h^{\rm U}(f;\theta)F(f;M_{\rm L}^{z},y),
\end{equation}
where $\theta$ represents the set of parameters defining an unlensed GW signal, $M_{\rm L}^{z}$ is the redshifted lens mass and $y$ is the dimensionless source position in source plane, and $F(f;M_{\rm L}^{z},y)$ is the amplification factor, which can be estimated by Equation~\eqref{eq:af}. Then we may estimate the Bayes factor $\mathcal{B}_{\rm U}^{\rm Micro}$, i.e., evidence ratio between microlensed and unlensed signal models, for all the events reported in GWTC catalog. Here we note that a high Bayes factor $\mathcal{B}_{\rm U}^{\rm Micro}$ itself does not necessarily mean a conclusive evidence of microlensing in an observed event, since the statistical fluctuation of $\mathcal{B}_{\rm U}^{\rm Micro}$ for unlensed signals may also be high enough and leads to false positives. For example, \citet{2021ApJ...923...14A} generated unlensed injections with Gaussian noise and recovered them with lensed templates, and they found that the typical statistical fluctuations of the resulting Bayes factors due to false alarm can be as high as $\log \mathcal{B}_{\rm U}^{\rm Micro} \sim 0.75$. \citet{2024ApJ...970..191A} estimated $\mathcal{B}_{\rm U}^{\rm Micro}$ for all the GW events in O3 and found that the distribution of $\log \mathcal{B}_{\rm U}^{\rm Micro}$ is centered at $0$ and the distribution does not deviate significantly with that from unlensed samples. Recently, \citet{2025arXiv251216347T} searched for the O4a data with similar procedures and found no compelling evidence for the presence of microlensing signatures in current GWTC data.

\section{Summary and Prospects}
\label{sec:sum}

GWs from astrophysical sources can be lensed by intervening masses, such as stars, primordial black holes, dark matter halos, galaxies, or galaxy clusters. When the GW wavelength is much smaller than the lens' characteristic scale (e.g., Einstein radius), lensing falls in the geometric-optics regime and can produce multiple time‑delayed images and overall magnification; when the wavelength is comparable to or larger than the lens' characteristic scale, wave‑optics effects produce frequency‑dependent magnification and phase shifts that manifest as interference fringes. These lensed signals are prime targets across the GW spectrum from high‑frequency ground‑based interferometers, low‑frequency space missions, and PTAs in the nanohertz band. Detecting them would uniquely probe the wave nature of GWs and yield constraints on CBC/SMBBH merger rates, the abundance and mass function of compact lenses (including primordial black holes), dark‑matter models, galaxy and cluster mass profiles, and cosmological parameters (most notably the Hubble constant) via measured time delays and magnifications.

For galaxy‑ and cluster‑scale lensing of high‑ and low‑frequency CBCs and SMBBH mergers in the geometric‑optics regime, detection prospects are encouraging. The typical optical depth for lensing by galaxies and clusters is on the order $10^{-4}-10^{-3}$, implying that one lensed event should appear once the cumulative number of detected GW events reaches $\sim 10^3-10^4$ even without considering magnification bias. The expected annual rate of ``detectable'' lensed CBCs/SMBBHs depends sensitively on detector sensitivity, source population models, and magnification bias. Extensive searches in the GWTC catalogs have produced several candidate lensing events, but none has been confirmed to date because false‑positive probabilities remain high for current LVK data. With the CBC sample already exceeding $300$ events and with magnification bias taken into account, a robust detection could plausibly occur in the near future, perhaps during O5 or with early runs of upgraded detectors. Third‑generation GW detectors are expected to detect on the order of tens to hundreds of lensed CBCs per year by the late 2030s, enabling a wide range of astrophysical and cosmological applications using these lensed events. For lensed GW events, sky localization may be dramatically improved by using an effective network of detectors that observe different lensed images from different space locations because of the detector's motion in the space, enabling the identification of their host galaxies and precise cosmological measurements. For millihertz space‑borne detectors, such as LISA, Taiji, and Tianqin, lensed SMBBH mergers are also expected to become detectable after these missions launch in the late 2030s. The expected detection rate is about $0.1-100$\,yr$^{-1}$, however, is highly uncertain and depends sensitively on the underlying SMBBH population model, particularly the seeding mechanism of SMBHs.

For wave-optics lensing of high-frequency GWs from CBCs by dark matter minihalo and compact mass (PBHs), the optical depth depends sensitively on both the lens population and source distribution (especially the source GW frequencies) because the lensing effects vary with the wavelength relative to the lens’s characteristic scale (Einstein radius). In this regime, the induced distortion of the GW waveform is typically small and multiple images are not produced, making detection difficult. Nevertheless, a subset of such lensed events may be observable with third generation ground-based GW detector networks and extreme-sensitive space missions such as DECIGO and BBO, offering valuable probes of the lens population and strong constraints on dark matter and PBHs.

For wave-optics lensing of nanohertz GW from inspiral SMBBHs lensed by galaxies, there could be a substantial lensing systems detectable by PTAs. Identifying these systems from PTA data alone is, however, challenging. For nearly circular SMBBH orbits, the signals are almost monochromatic, so the small, lens‑induced amplitude modulations and phase shifts are difficult to distinguish from unlensed signals or instrumental/systematic effects. By contrast, significantly eccentric binaries (with $e$ up to $\sim0.7$ or higher) emit GWs at multiple harmonics, producing richer, frequency‑dependent structure; in those cases the wave‑optics distortions induced by a lens are more likely to be separable from intrinsic source behavior. 

To search and identify lensed GW events in both geometric and wave‑optics, accurate waveform models that incorporate lensing effects are required in order to avoid biases and to search for fringes. For lensed events in the geometric-optics regime, it is required to include matched‑filter banks for lensed templates, morphological searches for spectral modulations, catalogs cross‑matching to find repeated events, careful parameter consistency checks and time‑delay matches. After the finding of some candidates for lensed events, it is important to note that distinguishing lensing from intrinsic waveform systematics, detector artifacts, or coincident astrophysical similarity is nontrivial and needs joint statistical approaches and even EM counterparts, when available. For lensed events in the wave-optics regime, the frequency‑dependent lensing signatures are typically subtle, but they may be recovered for rare, high‑S/N events in favorable geometries provided one uses detailed lens‑aware waveform models and matched‑filter searches. 

Once a population of lensed GW signals is firmly established, these events will be powerful tools for astrophysics and cosmology: they can constrain the event rate evolution of GW sources and the abundance and mass function of compact lenses (including PBHs), probe galaxy and cluster mass distributions, and serve as independent cosmological probes via measured time delays and magnifications. In particular, a statistical sample of lensed GW events will offer unique and independent constraints on cosmological parameters and on the nature of dark matter and PBHs, opportunities that are likely to mature over the next decade. Realizing this promise will require larger event samples, lens‑aware waveform models, robust statistical methods to control false positives, and coordinated multi‑messenger follow‑up to fully exploit the information encoded in lensed signals.

\section{Acknowledgment}
We thank Dr. Xiao Guo for providing the original data used to plot Figure~\ref{fig:dndz}. We also thank Dr. Yunfeng Chen, Dr. Yushan Xie, Yuchao Luo, and Yang Xie for helpful discussions and suggestions. This work is partly supported by the National Key program for Science and Technology Research and Development (grant nos. 2020YFC2201400 and 2022YFC2205201), the National Natural Science Foundation of China (grant no. 12273050), the Strategic Priority Program of the Chinese Academy of Sciences (grant no. XDB0550300), the National Astronomical Observatory of China (grant no. E4TG660101), and the Postdoctoral Fellowship Program of CPSF under Grant Number GZB20250735 (ZC).

\bibliographystyle{aasjournal}
%
%\bibliography{GWGLrev.bib}
\bibliography{bibtex1.bib}

\label{lastpage}

\end{document}